# Green modified function of the equation of the internal gravity waves in the stratum of the stratified medium with constant average flow

**Vitaly V. Bulatov, Yuriy V. Vladimirov**

**Institute for Problems in Mechanics
Russian Academy of Sciences
Pr.Vernadskogo 101-1, 117526 Moscow, Russia**
**bulatov@index-xx.ru**

Abstract.
*In the present paper construction of the modified function of Green equation for internal gravity waves in the stratum of the stratified medium at presence of constant average flows is considered, properties of the corresponding spectral problems, the modified eigenfunctions and eigenvalues are investigated. Usage of the modified function of Green equation can give in some physically interesting events more friendly representations of the solutions for the fields of the internal gravity waves, including the wave fields disturbed by the non-local disturbing bodies.*

Green equation for the internal gravity waves at presence of the average shift flows satisfies the equation

$$L\left(z, \frac{\partial}{\partial t}, \frac{\partial}{\partial x}, \frac{\partial}{\partial y}, \frac{\partial}{\partial z}\right)\Gamma = \delta(t)\delta(x)\delta(y)\delta(z-z'), \qquad (B1)$$

$$L = \frac{D^2}{Dt^2}\left(\frac{\partial^2}{\partial x^2} + \frac{\partial^2}{\partial y^2} + \frac{\partial^2}{\partial z^2}\right) - \frac{D}{Dt}\left(\frac{\partial^2 V_1}{\partial z^2}\frac{\partial}{\partial x} + \frac{\partial^2 V_2}{\partial z^2}\frac{\partial}{\partial y}\right) + N^2(z)\left(\frac{\partial^2}{\partial x^2} + \frac{\partial^2}{\partial y^2}\right)$$

$L$ - the operator of the internal waves in Boissinesq approximation, $N(z)$ is Brunt-Väisälä frequency; $V_1, V_2$ - the components of the current velocity $\mathbf{V} = \{V_1, V_2, 0\}$ on some horizon $z$ ($\sqrt{V_1^2 + V_2^2} = V$);

$$\frac{D}{Dt} = \frac{\partial}{\partial t} + V_1\frac{\partial}{\partial x} + V_2\frac{\partial}{\partial y}$$

Boundary and the initial conditions are taken in the form of

$$\Gamma = 0 \quad (z = 0, H); \qquad \Gamma \equiv 0, \quad t < 0. \qquad (B2)$$

By means of $\Gamma$ we construct the solution $U$ of the equation (B1) for the arbitrary right part of (B1): $Q = Q(t, x, y, z)$, depending on the particular statements of the problem

$$U = \int dt' \int dx' \int dy' \int dz' \, \Gamma(t-t', x-x', y-y', z-z') \, Q(t', x', y', z') \qquad (B3)$$

Green function for the internal waves by the Fourier transform on the variables $t, x, y$ is reduced to the following view

$$\Gamma = \frac{1}{(2\pi)^3} \int_{-\infty}^{\infty} d\lambda \int_{-\infty}^{\infty} d\mu \int_{-\infty+i\varepsilon}^{\infty+i\varepsilon} d\omega \, e^{i\lambda x + i\mu y - i\omega t} G(\omega, \lambda, \mu, z, z') \qquad (B4)$$

where G is the solution of the boundary problem
$$L_0 G = -\delta(z - z'); \quad G = 0, z = 0, H \tag{B5}$$
and $L_0$ is Taylor-Goldstein operator
$$L_0 = (\omega - f)^2 \frac{\partial^2}{\partial z^2} + \left\{ k^2 [N^2 - (\omega - f)^2] + \frac{\partial^2 f}{\partial z^2}(\omega - f) \right\}$$
$$f \equiv \lambda V_1(z) + \mu V_2(z), \qquad k^2 = \lambda^2 + \mu^2 \tag{B6}$$

The initial condition determines the bypass of the poles and cuts of G at integration of (B4) $(\varepsilon > 0, \varepsilon \to 0)$. The expression for G looks like

$$\Gamma = G(t, x, y, z, z') + \sum_{n=1}^{\infty} \frac{1}{2\pi^2} \operatorname{Im} \int_{-\infty}^{\infty} d\lambda \int_{-\infty}^{\infty} d\mu \, e^{i\lambda x + i\mu y - i\omega_n (\lambda, \mu) t} \frac{\varphi_n(z) \varphi_n(z')}{d_n (\omega_n - f(z'))^2} \tag{B7}$$

where $\varphi_n(z)$ and $\omega_n(\lambda, \mu)$ is the eigenfunctions and the eigenfrequencies of the operator $L_0$

$$d_n = \left. \frac{\partial \varphi_n}{\partial \omega} \frac{\partial \varphi_n}{\partial z} \right|_{z=H, \omega = \omega_n}$$

$G_m$ is the contribution of the continuous spectrum of the operator $L_0$.

The wave zone defined by (B7) at the great values of $|x|, |y|, |t|$ is limited by two curves - by the forward wave front and the rear wave front. At $V = \text{const}$ the contribution of the continuous spectrum and the rear wave front fade away.

Later we shall consider the Green function representations at $V = \text{const}$ in the modified form allowing more completely to determine the spatial structure of (B7) both at the great distances and in the immediate proximity to the sources of the internal gravity waves. It will be shown, that each mode of Green function consists of the sum of three terms, the first of which describes the internal waves propagating from the source, the second term describes the effects of the non-stationarity of the source localized in some vicinity around it, the third term describes the effects of the pushed away liquid (the internal jump) caused by the source. The analysis of the gained expressions for the constant and oscillating with the frequency $\Omega$ source $Q(t, x, y, z)$ will be conducted below, at that each of Green function items will be represented in the form of the single integral suitable for the analytical and numerical analysis.

The routine method of a solution of the non-uniform boundary problem of the (B5) type is expansion of G into the full set of the linearly-independent functions, being the solutions of the corresponding uniform problem.

At the same time it seems useful looking for the required solution to use the other full sets of the linearly-independent functions expansion, which in the certain sense meet the required statements of the problem and allows to gain some additional information about behavior of Green function. In the capacity of such set we shall take the solution of the following boundary problem

$$\frac{\partial^2 \psi_n(z, \chi)}{\partial z^2} + \chi \left[ \frac{N^2(z)}{v_n(\chi)} - 1 \right] \psi_n(z, \chi) = 0$$

$$\psi_n = 0, \quad z = 0, H \tag{B8}$$

$$\chi \in (-\infty, \infty), \quad v_n \in (-\infty, \infty)$$

At the positive values of $\chi$ the solution (B8) describes the vertical modes of oscillations of the particles in the stratified medium at lack of the flows with the values of the wave number square $\chi$ and the square of the frequency $v_n(\chi)$. At the negative $\chi$ as it will be shown below, (B8) describes the oscillations of the liquid in areas of the internal jumps. To distinguish the boundary problem (B8) from the normally used (in which $\chi$ is always $\chi > 0$), below we shall call (B8) as the modified boundary problem.

Let's introduce the following notation $\Lambda_n(\chi) = -\chi/v_n(\chi)$. At that (B8) shall to take the form of Sturm-Liouville boundary problem creating for any substantial $\chi$ the full set of the eigenfunctions $\psi_n(\chi,z)$ being inter-orthogonal in the space with the scalar product $(a,b)_{N^2} = \int_0^H N^2(z) a b^* dz$. The spectrum of the eigennumbers $\Lambda_n(\chi)$, $n = 1,2,K$ is substantial, discrete and contains no more than finite number of the negative values. The number $n$ of the eigenfunction $\psi_n(\chi,z)$ is determined by the number of its zero. To show it for any $\chi$ it is suitable to take advantage of the method of the phase functions. Let's represent the solution of the problem (B8) in the form of

$$\psi_n(\chi,z) = A_n(\chi,z)\left[\text{sh}(\chi^{1/2}z)\cos(B_n(\chi,z)) + \text{ch}(\chi^{1/2}z)\sin(B_n(\chi,z))\right] \quad (B9)$$

By analogy to the method of variation of parameters we shall superimpose on the function $A_n, B_n$ the following requirement

$$\frac{d}{dz}\psi_n = A_n\left[\cos B_n \frac{d}{dz}\text{sh}(\chi^{1/2}z) + \sin B_n \frac{d}{dz}\text{ch}(\chi^{1/2}z)\right] \quad (B10)$$

Substitution of (B9) in (B8) with the allowance for (B10) gives the equations for the functions $A_n, B_n$

$$A_n(\chi,z) = A_n(\chi,0)\exp\left[-\frac{\chi^{1/2}}{v_n(\chi)}\right] \times$$
$$\times \int_0^z dz N^2(z)\left[\cos B_n \text{sh}(\chi^{1/2}z) + \sin B_n \text{ch}(\chi^{1/2}z)\right] \times \left[\cos B_n \text{ch}(\chi^{1/2}z) - \sin B_n \text{sh}(\chi^{1/2}z)\right] \quad (B11)$$

$$\frac{dB_n}{dz} = \chi^{1/2}\frac{N^2(z)}{v_n(\chi)}\left[\cos B_n \text{sh}(\chi^{1/2}z) + \sin B_n \text{ch}(\chi^{1/2}z)\right]^2 \quad (B12)$$

Now we shall write (B9) in the form of

$$\psi_n = A_n \text{ch}^{1/2}(2\chi^{1/2}z)\sin\left[\text{mod}_{2\pi}(B_n) + \text{arctg th}(\chi^{1/2}z)\right] \quad (B13)$$

In view of the monotonicity $B_n$ (the sequent of (B12)) and the boundary conditions we come to the conclusion, that at $\chi > 0$ the number $n$ is determined by the quantity of zeros of the function $\psi_n$. At $\chi < 0$ it is enough in (B9) - (B13) to substitute $B_n$ with $iB_n$ and $\psi_n$ with $i\psi_n$. Then similarly to (B9) - (B13) we shall have

$$\psi_n = A_n\left[\text{ch}B_n \sin(|\chi|^{1/2}z) - \text{sh}B_n \cos(|\chi|^{1/2}z)\right] \quad (B14)$$

$$\frac{dB_n}{dz} = -|\chi|^{1/2}\frac{N^2(z)}{v_n(\chi)}\left[\text{ch}B_n \sin(|\chi|^{1/2}z) - \text{sh}B_n \cos(|\chi|^{1/2}z)\right]^2 \quad (B15)$$

$$\psi_n = A_n \text{ch}^{1/2}(2|\chi|^{1/2}z)\sin\left[\text{mod}_{2\pi}(|\chi|^{1/2}z) + \text{arctg th } B_n\right] \quad (B16)$$

From (B15), (B16) follows, that and at $\chi < 0$ the number n is determined by the quantity of zeros of the function $\psi_n$.

The behavior of $\psi_n(\chi, z)$ at variation of z (the distribution of zeros of the function $\psi_n$ on the axis z) is determined by the sign of the value $\chi\left[\dfrac{N^2(z)}{v_n(\chi)} - 1\right]$. At $\chi \geq 0$, $v_n \geq 0$ and $\chi \leq 0$, $\chi \leq 0$ zeros of the function $\psi_n$ are concentrated in the area of the axis z with the maximum value of the Brunt-Väisälä frequency $N(z)$. At $\chi < 0$, $v_n > 0$ zeros of the function $\psi_n$ are concentrated in the area of with the minimum value of $N(z)$. At $\chi > 0$, $v_n < 0$ there is no oscillating solutions.

Let' consider the dependence of $v_n(\chi)$ from $\chi$. From (B8) follow the ratios

$$\chi \int_0^H dz \left[\frac{N^2(z)}{v_n(\chi)} - 1\right] |\psi_n(z,\chi)|^2 = \int_0^H dz \left|\frac{d\psi_n(\chi,z)}{dz}\right|^2 \tag{B17}$$

$$\frac{dv_n(\chi)}{d\chi} = \frac{v_n^2(\chi)}{\chi} \frac{\int_0^H dz \left[\frac{N^2(z)}{v_n(\chi)} - 1\right] |\psi_n(\chi,z)|^2}{\int_0^H dz\, N^2(z) |\psi_n(z,\chi)|^2} \tag{B18}$$

From which follows the monotonicity $v_n(\chi)$, or the non-negativity $\dfrac{dv_n(\chi)}{d\chi}$ for any $\chi$. In view of correspondence of the quantity of zeros of the function $\psi_n$ to its number n we using (B8) determined five special points of the dispersion curves $v_n(\chi)$

1) $v_n(\chi) \to \max\limits_z N^2(z)$

$\chi \to +\infty$

2) $v_n(0) = 0$

3) $v_n(\chi) \to -\infty$

$\chi \to -\left(\dfrac{\pi n}{H}\right)^2 + 0$ \hspace{2em} (B19)

4) $v_n(\chi) \to +\infty$

$\chi \to -\left(\dfrac{\pi n}{H}\right)^2 - 0$

5) $v_n(\chi) \to \min\limits_z N(z)$

$\chi \to -\infty$

In particular, at $N^2(z) = N^2 = $ const we have

$$v_n(\chi) = \frac{\chi N^2}{\chi + \left(\dfrac{\pi n}{H}\right)^2} \tag{B20}$$

Comparing (B20) and (B17) we come to the conclusion, that for particular number n it is possible to introduce the effective thickness of the stratum of the stratified liquid, in which the internal wave is propagating, and effective Brunt-Väisälä frequency of this stratum

$$\tilde{H}_n(\chi) \equiv \left[ \frac{\int_0^H dz |\psi_n|^2}{\int_0^H dz \left| \frac{1}{\pi n} \frac{d\psi_n}{dz} \right|^2} \right]^{\frac{1}{2}} \tag{B21}$$

$$\tilde{N}_n(\chi) \equiv \left[ \frac{\int_0^H dz N^2(z) |\psi_n|^2}{\int_0^H dz |\psi_n|^2} \right]^{\frac{1}{2}} \tag{B22}$$

At that

$$\min_z N(z) \leq \tilde{N}_n(\chi) \leq \max_z N(z)$$

$$\tilde{H}_n\left(\frac{\pi n}{H}\right) = H$$

After that $\nu_n(\chi)$ will look like

$$\nu_n(\chi) = \frac{\chi \, \tilde{N}_n^2(\chi)}{\chi + \left(\frac{\pi n}{\tilde{H}_n(\chi)}\right)^2} \tag{B23}$$

And for $\frac{d\nu_n(\chi)}{d\chi}$ we shall have

$$\frac{d\nu_n(\chi)}{d\chi} = \frac{\nu_n(\chi)}{\chi}\left(1 - \frac{\nu_n(\chi)}{\tilde{N}_n^2(\chi)}\right) = \frac{\left(\frac{\pi n}{\tilde{H}_n(\chi)}\right)^2 \tilde{N}_n^2(\chi)}{\left[\chi + \left(\frac{\pi n}{\tilde{H}_n(\chi)}\right)^2\right]^2} \tag{B24}$$

The corresponding (B19)-(B24) graphs of the function $v_n(\chi)$ have the typical appearance shown on the Fig.B1

$$\Gamma = \sum_{n=1}^{\infty} \Gamma_n \tag{B26}$$

$$\Gamma_n = \frac{1}{(2\pi)^3} \int_{-\infty}^{\infty} d\omega \int_{-\infty}^{\infty} d\lambda \int_{-\infty}^{\infty} d\mu \, e^{i(\lambda x + \mu|y| - \omega t)} \frac{v_n(k^2)}{k^2 \left[(\omega + i\varepsilon - f)^2 - v_n(k^2)\right]} \frac{\psi_n(k^2, z)\psi_n^*(k^2, z)}{\int_0^H dz \, N^2(z) |\psi_n(k^2, z)|^2}$$

Then representing the solution of the boundary problem (B5) in the form of the set $\psi_n(\chi, z)$ for $\chi = k^2$ expansion

$$G = \sum_{n=1}^{\infty} p_n \psi_n(k^2, z) \tag{B25}$$

and solving $p_n$ with the help of convolution (5) with $\psi_n(k^2, z)$ we come to the following expression for Green function

$$\Gamma = \sum_{n=1}^{\infty} \Gamma_n \tag{B26}$$

$$\Gamma_n = \frac{1}{(2\pi)^3} \int_{-\infty}^{\infty} d\omega \int_{-\infty}^{\infty} d\lambda \int_{-\infty}^{\infty} d\mu \, e^{i(\lambda x + \mu|y| - \omega t)} \frac{v_n(k^2)}{k^2 \left[(\omega + i\varepsilon - f)^2 - v_n(k^2)\right]} \frac{\psi_n(k^2, z)\psi_n^*(k^2, z)}{\int_0^H dz \, N^2(z) |\psi_n(k^2, z)|^2} \tag{B27}$$

$$k^2 = \lambda^2 + \mu^2`$$

Considering, that the axis x is always directed against the current velocity $(f = -\lambda V)$, we gain the equation for the poles of the integrand (B27)

$$v_n = (\omega + i\varepsilon - f)^2 \tag{B28}$$

Or in view of the formula (B23)

$$\tilde{N}_n^2(k^2)(\lambda^2 + \mu^2) = (\omega + i\varepsilon + \lambda V)^2 \left[\lambda^2 + \mu^2 + \left(\frac{\pi n}{\tilde{H}_n(k^2)}\right)^2\right] \tag{B29}$$

The integration is most simply fulfilled with respect to $\omega$, for what close the contour of integration in the negative half-plane of the variable $\omega$ (complex) and gain after application of the residue theorem (for $t > 0$), which allows to take one of the integrals in (B26) - (B27).

$$\Gamma_n = -\frac{1}{(2\pi)^2} \int_{-\infty}^{\infty} d\lambda \int_{-\infty}^{\infty} d\mu \, e^{i(\lambda(x+tV) + \mu|y|)} \sin\left(t\sqrt{v_n(k^2)}\right) \frac{\sqrt{v_n(k^2)}}{k^2} \frac{\psi_n(k^2, z)\psi_n^*(k^2, z')}{\int_0^H dz \, N^2(z) |\psi_n(k^2, z)|^2} =$$

$$= -\frac{1}{2\pi} \int_{-\infty}^{\infty} dk \, J_0(kR) \sin\left(t\sqrt{v_n(k^2)}\right) \frac{\sqrt{v_n(k^2)}}{k^2} \frac{\psi_n(k^2, z)\psi_n^*(k^2, z')}{\int_0^H dz \, N^2(z) |\psi_n(k^2, z)|^2} \tag{B30}$$

$$R \equiv \left[(x+Vt)^2 + y^2\right]^{\frac{1}{2}}$$

where $J_0(kR)$ is Bessel function of the zero-order.

For $t<0$ the contour of integration is closed in the upper half plane, that result in $\Gamma_n = 0$ according to the initial condition of (B2). In the coordinate system moving together with the flow ($x \to x - Vt$) the expression (B30) possesses the circular symmetry and coincides with the expression for Green function gained in [47-69]. It is evident, that if values of $t, |x|, |y|$ are small, $\Gamma_n$ may be gained only by the numerical integration (B30)

In the practical applications, for example, in the problem about generation of the internal gravity waves by the pulsing sources, often it is suitable to use the spectral density $\tilde{\Gamma}_n$ determined by the formula

$$\Gamma_n = \frac{1}{2\pi} \int_{-\infty}^{\infty} \tilde{\Gamma}_n d\omega$$

In this case in (B27) it is necessary to conduct $\lambda$ or $\mu$ integration.

The $\lambda$ integration as applied to the problem about generation of the internal gravity waves by the source of the constant intensity ($\omega = 0$) was used in [47-69]. At that Green function was represented in the form of the eigenfunctions row of the general boundary problem (the modified problem (B8) coincides with the general problem at $\chi > 0$. As the result only the real poles $\lambda = \pm \lambda_n(\mu^2)$ are taken into consideration, that nevertheless does not change the asymptotics of Green function at the great values of $t, |x|, |y|$ determined only by the real poles.

It is easy to demonstrate, that usage of the modified boundary problem determines the presence besides the couple of the real poles also a couple of clearly imaginary poles $\lambda_n(\mu^2)$. Really, from (B29) follows (for the poles not shifted because of the component $i\varepsilon$ presence)

$$\lambda_n^2(\mu^2) = \frac{1}{2}\left[\frac{\tilde{N}_n(k^2)}{V^2} - \mu^2 - \left(\frac{\pi n}{\tilde{H}(k^2)}\right)^2\right] \pm \sqrt{\left[\frac{\tilde{N}_n(k^2)}{V^2} - \mu^2 - \left(\frac{\pi n}{\tilde{H}_n(k^2)}\right)^2\right]^2 + 4\mu^2 \frac{\tilde{N}_n(k^2)}{V^2}} \quad (B31)$$

$$k^2 = \mu^2 + \lambda_n^2(\mu^2)$$

As the values $\tilde{N}_n(k^2)$ and $\tilde{H}_n(k^2)$ by definition are real, then from (B31) follows the reality of $\lambda_n^2(\mu^2)$. We shall note, that if $N(z) \approx const$, then $\tilde{N}_k$ and $\tilde{H}_n$ practically do not depend on $k$ and (B31) directly determines the poles $\lambda_n(\mu^2)$. The shift of the real poles in the complex plane caused by the component $i\varepsilon$ is determined from expansion (B28) with respect to $\varepsilon$

$$\lambda_n(\mu^2)_\varepsilon = \lambda_n(\mu^2)_{\varepsilon=0} + i\varepsilon \frac{\partial \lambda_n(\mu^2)}{\partial \varepsilon} + \Lambda$$

where

$$\frac{\partial \lambda_n(\mu^2)}{\partial \varepsilon} = \frac{V}{\frac{\partial v_n(k^2)}{\partial k^2} - V^2} \tag{B32}$$

The spectral density $\tilde{\Gamma}_n$ at $\omega = 0$ is represented in the form of

$$\tilde{\Gamma}_n = \frac{1}{(2\pi)^2} \int_0^\infty d\mu \int_{-\infty}^\infty d\lambda\, e^{i(\lambda x + \mu|y|)} \frac{v_n(k^2)}{k^2[(\lambda V + i\varepsilon)^2 - v_n(k^2)]} \frac{\psi_n(k^2,z)\psi_n^*(k^2,z)}{\int_0^H dz\, N^2(z)|\psi_n(k^2,z)|^2} + K^* \tag{B33}$$

Here and below $K^*$ is the corresponding complex conjugate item. Later, considering the function $v_n(k^2)$ and the function

$$b_n(k^2) \equiv \frac{\psi_n(k^2,z)\psi_n^*(k^2,z)}{\int_0^H dz\, N^2(z)|\psi_n(k^2,z)|^2}$$

analytically prolonged into the area of the complex expression $\lambda$ ($k^2 = \lambda^2 + \mu^2$), we shall calculate the integral with respect to the variable $\lambda$ by means of the residue theorem. As a result, at the values of $x > 0$ we shall gain

$$\tilde{\Gamma}_n = -\frac{1}{2\pi}\int_0^\infty d\mu\, e^{i\mu|y|} \frac{\sin(\lambda_n(\mu^2)x)}{\lambda_n(\mu^2)} \frac{v_n(k^2)b_n(k^2)}{k^2\left[V^2 - \frac{\partial v_n(k^2)}{\partial k^2}\right]} + \frac{1}{2\pi}\int_0^\infty d\mu\, e^{i\mu|y|} \frac{e^{-|\lambda_n(\mu^2)|x}}{2|\lambda_n(\mu^2)|} \frac{v_n(k^2)b_n(k^2)}{k^2\left[V^2 - \frac{\partial v_n(k^2)}{\partial k^2}\right]} + K^*$$

$$\lambda_n^2(\mu^2) > 0 \qquad \frac{\partial v_n(k^2)}{\partial k^2} > V^2 \qquad\qquad \lambda_n^2(\mu^2) < 0 \tag{B34}$$

at the value of $x < 0$

$$\tilde{\Gamma}_n = -\frac{1}{2\pi}\int_0^\infty d\mu\, e^{i\mu|y|} \frac{\sin(\lambda_n(\mu^2)|x|)}{\lambda_n(\mu^2)} \frac{v_n(k^2)b_n(k^2)}{k^2\left[V^2 - \frac{\partial v_n(k^2)}{\partial k^2}\right]} + \frac{1}{2\pi}\int_0^\infty d\mu\, e^{i\mu|y|} \frac{e^{-|\lambda_n(\mu^2)||x|}}{2|\lambda_n(\mu^2)|} \frac{v_n(k^2)b_n(k^2)}{k^2\left[V^2 - \frac{\partial v_n(k^2)}{\partial k^2}\right]} + K^*$$

$$\lambda_n^2(\mu^2) > 0 \qquad \frac{\partial v_n(k^2)}{\partial k^2} < V^2 \qquad\qquad \lambda_n^2(\mu^2) < 0 \tag{B35}$$

where $k^2 = \mu^2 + \lambda_n^2(\mu^2)$ and $v_n(k^2) = V^2 \lambda_n^2(\mu^2)$.

From (B34), (B35) follows, that the integral with the positive values of $\lambda_n^2(\mu^2)$ determines the wave component of the source generated field

Allowing, that at $\lambda_n^2(\mu^2) \geq 0$ and ($k^2 = \mu^2 + \lambda_n^2(\mu^2) \geq 0$) the values $v_n(k^2)$ and $\frac{\partial v_n(k^2)}{\partial k^2}$ are limited.

$$v_n(k^2) < \max_z N^2(z)$$

$$\frac{\partial v_n(k^2)}{\partial k^2} \leq \left.\frac{\partial v_n(k^2)}{\partial k^2}\right|_{k^2=0} = \frac{\tilde{N}^2(0)\tilde{H}_n^2(0)}{\pi^2 n^2}$$

$$\lim_{\mu \to \infty} \frac{\partial v_n(k^2)}{\partial k^2} = 0$$,

We gain, that the wave field always exists in the field of $x < 0$. In the field of $x > 0$ there is only the final quantity of the modes of the internal waves, which numbers $n$ do not exceed the value $\frac{\tilde{N}_n(0)\tilde{H}_n(0)}{\pi V}$.

Later, for ease we shall consider the requirement $\frac{\tilde{N}_n(0)\tilde{H}_n(0)}{\pi V} < 1$ as always fulfilled. The asymptotics of the wave component $\tilde{\Gamma}_n$ ($\omega = 0$) at the great values of $|x|, |y|$ is determined by the standard methods and corresponds to the results gained in [47-69]. The integral with the negative $\lambda_n^2(\mu^2)$ in (B34), (B35) because of the presence of $e^{-|\lambda_n(\mu^2)||x|}$ describes the fast fading at the increase of $|x|$ component $\Gamma_n$. If the wave field disappears at absence of the stratification - $N(z) \equiv 0$ - (according to { B31) $\lambda_n^2(\mu^2)$ is the non-negative component only in the point $\mu = 0$), then as it will be shown later, the fast fading term of (B34), (B35) is different from zero, and due to that, describes the effects of the liquid expulsion by the source. We shall also note, that unlike the wave term, the integrand of the second integral in (B34), (B35) has the singularities in some areas of integration. Below, in the more general event of $\omega \neq 0$ it will be shown, that these singularities are removed at the calculation with the help of $\tilde{\Gamma}_n$ for the field of the internal waves from the lengthened along the y axis source – the well known problem of the energy infinity of the internal gravity waves generated by the point source

If $\omega \neq 0$, then the poles $\lambda_n(\omega, \mu^2)$ should be determined from the equation (B29), which already it is impossible to represent in the form of (B31) (the equation (B31) is gained by the solution of (B29), in which the values $\tilde{N}_n(k^2)$ and $\tilde{N}_n(k^2)$ formally are considered as not dependent on $k^2$). If by analogy with the conclusion from (B31) to consider $\tilde{N}_n(k^2)$ and $\tilde{H}_n(k^2)$ formally nondependent on $k^2$, then it is possible to gain the equations $\lambda_n(\omega, \mu^2)$ in the form of the solution of the equation of the fourth order (B26) with the help of Descartes-Euler method or Ferrari method allowing by virtue of the reality of $\tilde{N}_n(k^2)$ and $\tilde{H}_n(k^2)$ to discuss the behavior of the real and imaginary parts of the poles $\lambda_n(\omega, \mu^2)$. It is apparent, that at $\omega \neq 0$ the poles become complex, however because of the bulkiness of the gained expressions the analytical analysis of $\tilde{\Gamma}_n$ behavior is hampered.

The more acceptable expression of $\tilde{\Gamma}_n$ from the point of view of the analysis and calculation at any values of $x, y$ allows to gain the integration in (B27) with respect to the variable $\mu$. Before this it is suitable to make the substitution of the variables $\lambda \to \lambda - \omega/V$ and to represent $\tilde{\Gamma}_n$ in the form of

$$\tilde{\Gamma}_n = \frac{1}{(2\pi)^2} \int_0^\infty d\lambda \int_{-\infty}^\infty d\mu \, e^{i(\lambda x + \mu|y| - \omega(t + \frac{x}{V}))} \frac{\nu_n(k^2)}{k^2[(\lambda V + i\varepsilon)^2 - \nu_n(k^2)]} \frac{\psi_n(k^2, z)\psi_n^*(k^2, z')}{\tilde{N}_n^2(k^2) \int_0^z dz |\psi_n(k^2, z)|^2} + K^* \quad (B36)$$

$$k^2 = \left(\lambda - \frac{\omega}{V}\right)^2 + \mu^2$$

The poles $\mu_n(\lambda, \omega)$ of the integrand (B36) satisfy the equation

$$\nu_n(k^2) = (\lambda V + i\varepsilon)^2 \quad (B37)$$

Or with allowance for (B23)

$$\mu_n^2(\lambda, \omega) = -\left(\lambda - \frac{\omega}{V}\right)^2 + \left(\frac{\pi n}{\tilde{H}_n(k^2)}\right)^2 \left[\frac{\tilde{N}_n(k^2)}{\tilde{N}_n(k^2) - (\lambda V + i\varepsilon)^2} - 1\right] \quad (B38)$$

$$k^2 = \left(\lambda - \frac{\omega}{V}\right)^2 + \mu_n^2(\lambda, \omega)$$

From (B38) the follows the reality of squares of the non-shifted ($\varepsilon = 0$) poles $\mu_n^2(\lambda, \omega)$ for any values of $\lambda$ follows and $\omega$.

Owing to limitlessness of the function $\nu_n$ (one of the differences of the modified boundary problem from the normal problem) the equation (B37) is always solvable for $\mu_n^2(\lambda, \omega)$. At that in compliance with the behavior of $\nu_n$ (Fig. B1) the following situations can take place:

a) $\lambda < \frac{\min_z N(z)}{V}$ - there is one positive value of the parameter $\nu_n$ in (B37) and as the result of it - one pair of the poles $\mu_n(\lambda, \omega)$,

b) $\frac{\min_z N(z)}{V} < \lambda < \frac{\max_z N(z)}{V}$ - there is one positive and may be one negative value of the parameter $\nu_n$ in (B37) and accordingly two pairs (pure real and pure imaginary) poles $\mu_n(\lambda, \omega)$ (and both pairs may be pure imaginary $\mu_n$),

c) $\frac{\max_z N(z)}{V} < \lambda$ - there is one negative value of the parameter $\nu_n$ in (B37), that is one [air of the pure imaginary poles $\mu_n(\lambda, \omega)$.

We should note, that if Brunt-Väisälä frequency $N(z)$ weakly depends on $z$, then the values of $\tilde{N}_n(k^2)$ and $\tilde{H}_n(k^2)$ as the first approximation do not depend on $k^2$, and (B38) directly determines $\mu_n(\lambda, \omega)$.

The Fig. B2, B3, B4 show the graphs of $\mu_n^2(\lambda, \omega)$ at $\omega = 0$, $\omega > 0$ and $\omega < 0$. Digits I, II, III mark the sections of the curves $\mu_n^2(\lambda, \omega)$ of the definite sign of $\mu_n^2$. At that any $\lambda$ and $\omega$ meets

$$\left|\mu_n^{II}(\lambda, \omega)\right| < \frac{\omega}{V}; \qquad\qquad \left|\mu_n^{III}(\lambda, \omega)\right| > \frac{\pi n}{\tilde{H}_n(\infty)} \quad (B39)$$

Considering the functions $v_n(k^2)$ and $\tilde{b}_n(k^2) \equiv \dfrac{\psi_n(k^2, z)\psi_n^*(k^2, z')}{\int_0^h dz |\psi_n(k^2, z)|^2}$

as analytically prolonged in the area of the complex $\mu$ we shall take the integral with respect to $\mu$ in (B36) by means of the residue theorem. Allowing for the shift of the poles from the real axis, we close the contour of integration in the upper half plane. Using the formulas (B23), (B24), and also (B37), (B38) we shall gain for $\tilde{\Gamma}_n$

$$\tilde{\Gamma}_n = (\tilde{\Gamma}_n^I + \tilde{\Gamma}_n^{II} + \tilde{\Gamma}_n^{III})e^{-i\omega(t+\frac{x}{V})} \tag{B40}$$

where

I) $\mu_n^2 = (\mu_n^I)^2 \geq 0$

$$\tilde{\Gamma}_n^I = \frac{1}{4\pi} \int_0^{\max N(z)/V} d\lambda\, e^{i(\lambda x + |\mu_n(\lambda,\omega)||y| - \frac{\pi}{2})} \frac{\tilde{b}_n(k^2)}{|\mu_n^I(\lambda,\omega)|[\lambda^2 V^2 - \tilde{N}_n^2(k^2)]} + K^* \tag{B41}$$

II) $\mu_n^2 = (\mu_n^I)^2 \geq 0$

$$\tilde{\Gamma}_n^{II} = \frac{1}{4\pi} \int_{\min N(z)/V}^{\infty} d\lambda\, e^{i\lambda x} e^{-|\mu_n^{II}(\lambda,\omega)||y|} \frac{\tilde{b}_n(k^2)}{|\mu_n^{II}(\lambda,\omega)|[\lambda^2 V^2 - \tilde{N}_n^2(k^2)]} + K^* \tag{B42}$$

III) $\mu_n^2 = (\mu_n^I)^2 \geq 0$

$$\tilde{\Gamma}_n^{III} = \frac{1}{4\pi} \int_0^{\max N(z)/V} d\lambda\, e^{i\lambda x} e^{-|\mu_n^{III}(\lambda,\omega)||y|} \frac{\tilde{b}_n(k^2)}{|\mu_n^{III}(\lambda,\omega)|[\lambda^2 V^2 - \tilde{N}_n^2(k^2)]} + K^* \tag{B43}$$

Here $k^2 = \lambda^2 + \mu_n^2(\lambda,\omega)$

The integrands in (B41) - (B43) have a singularity in the points of $\lambda$, where $|\mu_n(\lambda,\omega)| = 0$ also the essential singularity in the points $\lambda^2 V^2 = \tilde{N}_n(k^2)$. If the source Q in (B3) is punctual, then (B40) describes the field of the waves formed by it and the essential singularity in integrals remains. If Q is the source lengthened along axis y, then at the calculation (B3) may be recorded as

$$\int_{-\infty}^{\infty} dy' \Gamma_n(y-y') Q(y') = \sum_\lambda Q(y_\lambda) \overline{\Gamma}_n(y - y_\lambda, \Delta y_\lambda)$$

where $\overline{\Gamma}_n(y, \Delta y_\lambda) = \dfrac{1}{2\Delta y_\lambda} \int_{y_\lambda - \Delta y_\lambda}^{y_\lambda + \Delta y_\lambda} dy' \tilde{\Gamma}_n(y)$ \hfill (B44)

$$2\Delta y_\lambda = y_{\lambda+1} - y_\lambda \tag{B45}$$

(B46)

Expression for the "averaged" function $\overline{\Gamma}_n$ differs by introduction in the integrands of the normalizing multipliers of the following type

$$\frac{\sin(|\mu_n(\lambda,\omega)|\Delta y_\lambda)}{|\mu_n(\lambda,\omega)|} \quad (B46)$$

As a result in the denominator of (B41) - (B43) there is the product $\frac{\sin(|\mu_n(\lambda,\omega)|\Delta y_\lambda)}{|\mu_n(\lambda,\omega)|}$ of (B46), that with allowance for (B38) removes the essential singularity.

In the case of the source non-lengthened along axis y such regularization is not possible.

Another singularity arising at $|\mu_n(\lambda,\omega)| \to 0$ is removed with the help of the requirement of generation, according to which Green function should to convert into zero at $|y| \to \infty$. The contribution of this singularity to the integral apparently does not depend on y and for its removal it is enough to subtract from the Internal its value in any other point. As for the calculation of $\tilde{\Gamma}_n$ it should be $\tilde{\Gamma}_n(|y|\to\infty)=0$, then it is possible to use the step by step calculation (beginning from $y \gg 1$)

$$\tilde{\Gamma}_n(|y|-\Delta y) = \tilde{\Gamma}_n(|y|+\Delta y) + \tilde{\tilde{\Gamma}}_n(|y|,\Delta y) \quad (B47)$$

The value of $\tilde{\tilde{\Gamma}}_n$ is determined by the ratio of (B47) identically and the formula for it differs from $\tilde{\Gamma}_n$ by introduction in the integrand of the multiplier $-2i\sin(|\mu_n|\Delta y)$ removing the singularity at $\mu_n = 0$

The described procedure of the regularization also improves convergence of the partial sums at calculation of Green function as the sum with respect to n (because at $n \to \infty$ the value of $\mu_n$ is proportional to n). The regularization of expressions (B34), (B35) for $\tilde{\Gamma}_n$ is similarly conducted at $\omega = 0$. For this purpose it is necessary to transfer from the variable $\mu$ in the integral to other variable T determined for the corresponding sections of the monotonicity of the curve $|\lambda_n(\mu^2)|$ by the formula $\mu = \mu_n(T^2)$ from which we gain

$$\lambda_n^2(\mu_n^2(S^2)) = T^2 \quad (B48)$$

$$d\mu \frac{v_n(k^2)/k^2}{|\lambda_n(\mu^2)|\left[V^2 - \frac{\partial v_n(k^2)}{\partial k^2}\right]} = dT \frac{\tilde{N}_n^2(k^2)}{|\mu_n(T^2)|\left[\tilde{N}_n^2(k^2) - V^2 T^2\right]} \quad (B49)$$

$$\frac{d|\lambda_n(\mu^2)|}{d\mu} = \frac{\mu \frac{\partial v_n(k^2)}{\partial k^2}}{|\lambda_n(\mu^2)|\left[V^2 - \frac{\partial v_n(k^2)}{\partial k^2}\right]} \quad (B50)$$

That reduces singularities of the integrands (B34), (B35) to the singularities of the integrands (B41) - (B43). The monotonicity $|\lambda_n(\mu^2)|$ on the sections of the integration of (B34), (B35) is proved by differentiation of (B28) with respect tp $\mu$

$$\frac{d|\lambda_n(\mu^2)|}{d\mu} = \frac{\mu \frac{\partial v_n(k^2)}{\partial k^2}}{|\lambda_n(\mu^2)|\left[V^2 - \frac{\partial v_n(k^2)}{\partial k^2}\right]} \tag{B50}$$

Whence in view of the formulas (B23), (B24) we have the signdefiniteness of $d\lambda_n / d\mu$ on the given intervals. Hereinafter the presence of the indicated regularization everywhere is meant.

If the stratification of the liquid is absent ($N(z) \equiv 0$), then according to {(B41) - (B43) the terms of $\tilde{\Gamma}_n^I$ and $\tilde{\Gamma}_n^{III}$ are fading, and $\tilde{\Gamma}_n^{II}$ becomes

$$\tilde{\Gamma}_n^{II} = \frac{1}{4\pi}\int_0^\infty d\lambda\, e^{i\lambda x} e^{-|y|\sqrt{(\lambda-\frac{\omega}{V})^2+(\frac{\pi n}{H})^2}} \frac{b_n(k^2)}{\lambda^2 V^2 \sqrt{(\lambda-\frac{\omega}{V})^2+(\frac{\pi n}{H})^2}} + K^* \tag{B51}$$

At the calculation of the field W of the vertical component of the velocity from the point source, as it is known, it is necessary to differentiate Green function with respect to z and twice to apply the operator $\frac{D}{Dt} = \frac{\partial}{\partial t} - V\frac{\partial}{\partial x}$. As a result from the denominator of (B51) the value $\lambda^2 V^2$ will disappear, and with consideration of the regularizing procedure of (B47) the expression for the field W will look like

$$W = \frac{1}{2\pi}\int_0^\infty d\lambda\, e^{i\lambda x} e^{-|y|\sqrt{(\lambda-\frac{\omega}{V})^2+(\frac{\pi n}{H})^2}} \frac{sh\left(\Delta y \sqrt{(\lambda-\frac{\omega}{V})^2+(\frac{\pi n}{H})^2}\right)}{\sqrt{(\lambda-\frac{\omega}{V})^2+(\frac{\pi n}{H})^2}} \frac{d\tilde{b}_n(k^2,z)}{dz} + K^* \tag{B52}$$

For the greater obviousness and convenience of the analysis of the formulas (B41) - (B43) we shall conduct in integrals (B42), (B43) the substitution of the variable $\lambda$ with the variable k by the formula

$$\lambda = \lambda(k) = \frac{\sqrt{v_n(k^2)}}{V} \tag{B53}$$

Realization of such a substitution follows from properties of the function $v_n(\chi)$. At that we shall gain

$$\mu_n^2(\lambda(k),\omega) = k^2 - \left(\omega - \sqrt{v_n(k^2)}\right)^2 \frac{1}{V^2} \tag{B54}$$

And accordingly for $\tilde{\Gamma}_n^I$ and $\tilde{\Gamma}_n^{III}$ we shall gain

$$\tilde{\Gamma}_n^I = -\frac{1}{4\pi}\int_0^\infty dk\, e^{i\left(\sqrt{\nu_n(k^2)}\frac{x}{V}+\frac{|y|}{V}\sqrt{k^2V^2-\left(\omega-\sqrt{\nu_n(k^2)}\right)^2}+\frac{\pi}{2}\right)} \frac{\sqrt{\nu_n(k^2)}}{k} \frac{\tilde{b}_n(k^2)}{\sqrt{k^2V^2-\left(\omega-\sqrt{\nu_n(k^2)}\right)^2}} + K^* \qquad (B55)$$

At $\quad k^2 > \dfrac{\left(\omega-\sqrt{\nu_n(k^2)}\right)^2}{V^2}$, and

$$\tilde{\Gamma}_n^{III} = -\frac{1}{4\pi}\int_0^\infty dk\, e^{i\sqrt{\nu_n(k^2)}\frac{x}{V}} e^{-\frac{|y|}{V}\sqrt{\left(\omega-\sqrt{\nu_n(k^2)}\right)^2-k^2V^2}} \frac{\sqrt{\nu_n(k^2)}}{k} \frac{\tilde{b}_n(k^2)}{\sqrt{\left(\omega-\sqrt{\nu_n(k^2)}\right)^2-k^2V^2}} + K^* \qquad (B56)$$

At $\quad k^2 > \dfrac{\left(\omega-\sqrt{\nu_n(k^2)}\right)^2}{V^2}$

We shall note, that $\tilde{\Gamma}_n^I + \tilde{\Gamma}_n^{III}$ may be recorded as one integral with respect to k from 0 up to $\infty$. By analogy with (B53) - (B56) we shall conduct in the integral (B41) substitution of the variable $\lambda$ with the variable k by the formula

$$\lambda = \tilde{\lambda}(k) = \frac{\sqrt{\nu_n(-k^2)}}{V} \qquad (B57)$$

That also ensures variation $\lambda$ in the necessary limits. At that we shall gain

$$\mu_n^2(\tilde{\lambda}(k),\omega) = -\left[k^2 - \left(\omega-\sqrt{\nu_n(-k^2)}\right)^2 \frac{1}{V^2}\right] \qquad (B58)$$

And accordingly for $\tilde{\Gamma}_n^{II}$

$$\tilde{\Gamma}_n^{II} = \frac{1}{4\pi}\int_{\frac{\pi m}{H}}^\infty dk\, e^{i\sqrt{\nu_n(-k^2)}\frac{x}{V}} e^{-\frac{|y|}{V}\sqrt{\left(\omega-\sqrt{\nu_n(-k^2)}\right)^2+k^2V^2}} \frac{\sqrt{\nu_n(-k^2)}}{k} \frac{\tilde{b}_n(-k^2)}{\sqrt{\left(\omega-\sqrt{\nu_n(-k^2)}\right)^2+k^2V^2}} + K^* \qquad (B59)$$

Formulas (B55), (B56.) (B59) and (B40) evidently demonstrate dependence of $\tilde{\Gamma}_n$ on coordinates t, x, y, z and characteristics of stratification $\nu_n(\chi)$ and $\psi_n(\chi,z)$.

Value $\tilde{\Gamma}_n^I$ determines the wave component of Green function, which asymptotics is determined at the great values of $|x|,|y|$ by the points of the stationary phase satisfying to the equation

$$\frac{x}{V}\frac{\partial\sqrt{v_n(k^2)}}{\partial k^2} + \frac{|y|}{V}\frac{kV^2 + \left(-\sqrt{v_n(k^2)} + \omega\right)\frac{\partial\sqrt{v_n(k^2)}}{\partial k^2}}{\sqrt{k^2V^2 - \left(\omega - \sqrt{v_n(k^2)}\right)^2}} = 0 \tag{B60}$$

The lines of the constant phase describing the wave fronts are set at that by the equation

$$\sqrt{v_n(k^2)}\frac{x}{V} + \frac{y}{V}\sqrt{k^2V^2 - \left(\omega - \sqrt{v_n(k^2)}\right)^2} - \omega\left(t + \frac{x}{V}\right) = \Phi = \text{const} \tag{B61}$$

In (B61) it is necessary to substitute the value of k gained from (B60), however the more convenient method of receiving the lines of the constant phase consists in the representation of these lines in the parametric form $x = x(k), y = y(k)$. At that the functions $x(k), y(k)$ are gained by the solution of the system of two equations (B60), (B61), where k it is considered as the parameter.

Direct calculation gives for the lines of the constant phase the pattern of the waves, close to the standard Kelvin wave wedge. The wave pattern is symmetrical with respect to the axis x and the pattern boundary can be determined from (B60)

$$\max\frac{|y|}{x} = \max_k\left\{\frac{\frac{\partial\sqrt{v_n(k^2)}}{\partial k^2}\sqrt{V^2 - \left(\frac{\sqrt{v_n(k^2)}}{k} - \frac{\omega}{V}\right)^2}}{V^2 - \frac{\sqrt{v_n(k^2)}}{k}\frac{\partial\sqrt{v_n(k^2)}}{\partial k^2} + \frac{\omega}{k}\frac{\partial\sqrt{v_n(k^2)}}{\partial k^2}}\right\} \tag{B62}$$

If to consider, that $\dfrac{\partial\sqrt{v_n(k^2)}}{\partial k} = C_n^g(k)$, $\dfrac{\sqrt{v_n(k^2)}}{k} = C_n^f(k)$

is the group and phase velocities of the plane internal wave with a wave number k it is receivable (46) in the form of

$$\max\frac{|y|}{x} = \max_k\left\{\frac{C_n^g(k)\sqrt{V^2 - \left(C_n^f(k) - \frac{\omega}{k}\right)^2}}{V^2 - C_n^f(k)C_n^g(k) + \frac{\omega}{k}C_n^g(k)}\right\} \tag{B63}$$

In the system of coordinates connected with the source, the velocity of relocation of the wave zone S in the tangential direction is determined by the expression ($\theta$ - the angle of the wave wedge opening)

$$S(\omega) = V \sin \theta_{max} = \frac{V}{\sqrt{1 + ctg^2 \theta_{max}}} = \frac{V}{\sqrt{1 + \frac{1}{(\max(|y|/x))^2}}} =$$

$$= \max_k \left\{ \frac{V}{\sqrt{1 + \frac{x^2}{y^2}}} \right\} = \max_k \left\{ \frac{C_n^g(k) \sqrt{V^2 - \left(C_n^f - \frac{\omega}{k}\right)^2}}{\sqrt{(C_n^g)^2 + V^2 + 2\left(\frac{\omega}{k} - C_n^f(k)\right) C_n^g(k)}} \right\} \quad \text{(B64)}$$

From (B64) follows, that at $\omega = 0$ the velocity of relocation of the wave-front is equal to

$$S(0) = \frac{C_n^g(0) \sqrt{V^2 - (C_n^f(0))^2}}{\sqrt{V^2 + C_n^g(0)(C_n^g(0) - C_n^f(0))}} =$$

$$= \frac{C_n^g(0) \sqrt{V^2 - (C_n^f(0))^2}}{\sqrt{V^2 - (C_n^f(0))^2 + (C_n^f(0) - C_n^g(0))^2 + C_n^g(0)C_n^f(0)}} \quad \text{(B65)}$$

Or considering, that $C_n^g(0) = C_n^f(0)$

$$S(0) = C_n^g(0) \sqrt{1 - \frac{(C_n^f(0))^2}{V^2}} \quad \text{(B66)}$$

that is the velocity of relocation of the wave-front set does not exceed the maximum possible velocity of the internal gravity waves. If $\omega \neq 0$, then S apparently will be different for $\omega = +|\omega_0|$ and $\omega = -|\omega_0|$, the situation met at the construction of the field of the internal waves from oscillating sources, and, for example, for an event of the oscillating source it will lead to formation of two wave zones superimposed on each other, which boundaries do not coincide. At that we shall mark, that the wave in the last case is modulated according to (B40) frequency of $\omega_0$ oscillation.

The item $\tilde{\Gamma}_n^{II}$ (B59) is quickly fading at the increase of $|y|$. Its singularity is the behavior in the vertical direction as the function $\psi_n(\chi, z)$ at $\chi = -k^2$ is localized in the stratums with the minimal values of Brunt-Väisälä frequency. Considering, that $\tilde{\Gamma}_n^{II}$ at the absence of stratification ($N(z) \equiv 0$) does not transform in zero, it is possible to draw the conclusion, that this item determines the effects of the flow streamlining of the source (the internal jumps).

The item $\tilde{\Gamma}_n^{III}$ (B56) is different from zero only in the case of $\omega \neq 0$ (because as it has been noted, we suppose $V^2 > (C_n^f(k))^2 = \frac{v_n(k^2)}{k^2}$) and thus determines the effects of the nonstationarity of the sources of the internal waves. The asymptotics $\tilde{\Gamma}_n^{III}$ at the great values of $|x|$, $|y|$ wanes less fast, than $\tilde{\Gamma}_n^{II}$, that is stipulated by behavior of the index of the exponential curve in the integrand (B56) in the

complex plane of the variable k .In this case the asymptotics can be evaluated by the saddle point approximation.

## References

1.  Abduhl, K. M. N. 1979 Effects of wave drag on submerged bodies. *Houille blanche*, 8, 465-470.

2.  Abraham, G. 1963 Model study of water gravity waves generated by a moving circular low-pressure area. *J. Geophys. Res*., 68(8), 2185-2210.

3.  Abramowitz, M. and Stegun, I. A. (eds.) 1965 Handbook of Mathematical Functions. Dover.

4.  Aksenov, A. V., Mozhaev, V. V., Skorovarov, V. E. and Sheronov, A. A. 1989 Stratified flow over a cylinder at low values of the internal Froude number. *Fluid Dyn*. 24, 639-642.

5.  Akylas, T. R. 1984 On the excitation of long nonlinear water waves by a moving pressure distribution. *J. Fluid Mech*., 141, 455-466.

6.  Akylas, T. R. 1984 On the excitation of nonlinear water waves by a moving pressure distribution oscillating at resonant frequency. *Phys. Fluids*, 27(12), 2803-2807.

7.  Amen, R. and Maxworthy, T. 1980 The gravitational collapse of a mixed region into a linearly stratified fluid. *J. Fluid Mech*., 96(1), 65-80.

8.  Andersen, P. and Wizhou, H. 1986 On the calculation of two-dimensional added mass and damping coefficients by simple Green's function technique. *Ocean Eng*., 12(5), 425-451.

9.  Andrews, D.G., Holton, J.R. and C.B. Leovy 1987 Middle atmosphere dynamics. International Geophysics Series, Academic Press.

10. Appleby, J. C. and Crighton, D. G. 1986 Non-Boussinesq effects in the diffraction of internal waves from an oscillating cylinder. *Q. J. Mech. Appl. Maths* 39, 209-231.

11. Appleby, J. C. and Crighton, D. G. 1987 Internal gravity waves generated by oscillations of a sphere. *J. Fluid Mech*. 183, 139-150.

12. Babich, V.M. and Buldyrev, V.S. 1991 Short-Wavelenght Difraction Theory - Asymptotic Methods. *Vol.4 of Springer Series on Wave Phenomena* (Springer-Verlag)

13. Baines, P. G. 1971 The reflexion of internal/inertial waves from bumpy surfaces. *J. Fluid Mech*. 46, 273-291.

14. Baines, P. G. 1979 Observations of stratified flow over two-dimensional obstacles in fluid of finite depth. T*ellus*, 31(4), 351-371.

15. Baines, P. G. 1979 Observations of stratified flow past three-dimensional barriers. *J. Geophys. Res*., 84(12), 7834-7838.

16. Baines, P. G. 1987 Upstream blocking and airflow over mountains. *Ann. Rev. Fluid Mech*. 19, 75-97.

17. *Baines, P. G. and Grimshaw, R. H. J. 1979 Stratified flow over finite obstacles with weak stratification.* Geophys. Astrophys. Fluid Dyn. *13, 317-334.*

18. *Baines, P. G. and Hoinka, K. P. 1985 Stratified flow over two-dimensional topography in fluid of infinite depth: a laboratory simulation.* J. Atmos. Sci. *42, 1614-1630.*

19. Barber, N.F. 1963 The directional resolving power of an array of wave detectors. *Ocean wave spectra*. N.Y.: Engelwood Cliffs, Prentice Hall, 137-150.

20. Barcilon, V. and Bleistein, N. 1969 Scattering of inertial waves in a rotating fluid. *Stud. Appl. Maths* 48, 91-104.

21. Barnard, B. J. S. and Pritchard, W. G. 1975 The motion generated by a body moving through a stratified fluid at large Richardson number. *J. Fluid Mech*., 71(1), 43-64.

22. Basak, P. C. 1975 Surface waves in deep water due to arbitrary periodic surface pressure. *Meccanica*, 10(1), 42-48.

23. Batchelor, G. K. 1967 An Introduction to Fluid Dynamics. Cambridge University Press.

24. Bell, T. M. 1975 Lee waves in stratified flows with simple harmonic time dependence. *J. Fluid Mech*., 67(4), 705-722.

25. Belotserkovskii, O. M., Belotserkovskii, S. O., Gushchin, V. A., Morozov, E. N., Onufriev, A. T. and Ul'yanov, S. A. 1984 Numerical and experimental modeling of internal gravity waves during the motion of a body in a stratified liquid. *Sov. Phys. Dokl*. 29, 884-886.

26. Bleistein, N. 1966 Uniform asymptotic expansions of integrals with stationary point near algebraic singularity. *Comm. Pure Appl. Maths*. 19, 353-370.

27. Bleistein, N. 1984 Mathematical Methods for Wave Phenomena. Academic.

28. Bleishtein, N. and Handelsman, R. A. 1986 Asymptotic Expansions of Integrals. Dover.

29. Blumen, W. and McGregor, C. D. 1976 Wave drag by three-dimensional mountain lee-waves in nonplanar shear flow. *Tellus* 28, 287-298.

30. Bonneton, P. and Chomaz, J.-M. 1992 Instabilites du sillage genere par une sphere. *C. R. Acad. Sci*. *Paris* II 314, 1001-1006.

31. Bonneton, P., Chomaz, J.-M. and Hopfinger, E. J. 1993 Internal waves produced by the turbulent wake of a sphere moving horizontally in a stratified fluid. *J. Fluid Mech*. (in press).

32. Borovikov, V.A., Vladimirov, Y.V.and Kelbert, M.Y. 1984 Field of internal gravity waves excited by localized sources. *Izv.Akad.Nauk SSSR. Atm.Oceanic Physics*, 6, 494-498.

33. Borovikov, V.A.and Bulatov, V.V. 1986 On the applicability of limits of asymptotic formulas for the field of internal waves excited by a moving source. *Izv.Akad.Nauk SSSR. Atm.Oceanic Physics*, 6, 508-510.

34. Borovikov, V.A., Bulatov, V.V.and Kelbert, M.Y. 1988 Intermediate asymptotic behavior of the far field of internal waves in a layer of stratified fluid lying on a homogeneous layer. *Izv.Akad.Nauk SSSR. Fluid Dynamics*, 3, 453-456.

35. Borovikov, V.A., Bulatov, V.V., Vladimirov, Y.V. and Levchenko, E.S. 1989 Internal wave field generated by a source rest in a moving stratified fluid. *Appl.Mech.Techn.Phys.*, 4, 563-566.

36. Borovikov,V.A., Bulatov, V.V., Vladimirov, Y.V., et.al. 1994 Processing and analysis of internal wave measurements in the shelf region of Western Sahara. *Oceanology*, 4, 457-460.

37. Borovikov, V.A., Bulatov V.V. and Vladimirov, Y.V. 1995 Internal gravity waves excited by a body moving in a stratified fluid. *Fluid Dyn. Res.*, 5, 325-336.

38. Borovikov, V.A., Bulatov, V.V., Gilman, M.A. and Vladimirov Y.V. 1998 Internal gravity waves excited by a body moving an a stratified fluid. *Preprints of Twenty-Second Symposium of Naval Hydrodynamics*, Washington, D.C., U.S.A., 92-99.

39. Borovikov, V.A., Bulatov, V.V. and Morozov E.G. 1998 Localization of tidal internal waves in the tropical Atlantic: nonspectral and spectral approaches. *Proceeding of the Fourth International Conference on Mathematical and Numerical Aspects of Wave Propagation.* SIAM, Golden, Colorado, U.S.A., 787-789.

40. Borovikov, V.A., Bulatov, V.V., Morozov, E.G. and Tamaya, R.G. 1998 Nonspectral and spectral approach to the study of tidal internal wave propagation in the ocean (by the example of the Mesopolygon experiment). *Oceanology*, 3, 307-311.

41. Boyer, D. L., Davies, P. A., Fernando, H. J. S. and Zhang, X. 1989 Linearly stratified flow past a horizontal circular cylinder. *Phil. Trans. R. Soc. Lond*. A 328, 501-528.

42. Brekhovskikh, L. and Goncharov, V. 1985 Mechanics of Continua and Wave Dynamics. Springer.

43. Bretherton, F. P. 1967 The time-dependent motion due to a cylinder moving in an unbounded rotating or stratified fluid. *J. Fluid Mech*. 28, 545-570.

44. Brighton, P. W. M. 1978 Strongly stratified flow past three-dimensional obstacles. *Q. J. R. Met. Soc*. 104, 289-307.

45. Broutman, D., Macaskill, C., McIntyre, M.E. and Rottman, J.W. 1997 On Doppler-spreading models of internal waves *Geophys.Res.Lett.,* 24, 2813-2816.

46. Buhler, O. and McIntyre, M.E. 1998 On non-dissipative wave mean interactions in the atmosphere or ocean. *J.Fluid Mech.*, 354, 301-343.

47. Bulatov, V.V. 1991 Steady motion of stratified fluid over a rough bottom. *J.Appl.Mech.Techn.Phys.*, 5, 678-683.

48. Bulatov, V.V. and Vladimirov, Y.V. 1989 Propagation of Airy and Frenel internal waves in inhomogeneous media *Morsk.Gigrofiz.Zhurn.*, 6, 14-19(In Russian).

49. Bulatov, V.V. and Vladimirov, Y.V. 1990 Propagation of Airy and Frenel internal waves in unsteady media *Morsk.Gidrofiz.Zhurn.*, 5, 13-18(In Russian).

50. Bulatov, V.V. and Vladimirov, Y.V. 1990 Internal gravity waves in an inhomogeneous ocean. *Advanced Experimental Technics and CAE Methods in Ship Hydro- and Aerodynamics*. Varna, 1, 46/1-46/3.

51. Bulatov, V.V. and Vladimirov, Y.V. 1991 Internal gravity waves in an inhomogeneous ocean. *Mathematical and Numerical Aspects of Wave Propagation Phenomena*. SIAM, Philadelphia, USA, 700-702.

52. Bulatov, V.V. and Vladimirov, Y.V. 1991 Near field of internal waves excited by a source in a moving stratified liquid. *Appl.Mech.Techn.Phys.*, 1, 21-25.

53. Bulatov, V.V. and Vladimirov, Y.V. 1991 Internal gravity waves excited by a source in stratified horizontally inhomogeneous media. *Izv.Akad.Nauk SSSR . Fluid Dynamics*, 1, 102-105.

54. Bulatov, V.V. and Vladimirov, Y.V. 1993 Internal gravity waves from a body moving in a stratified ocean. *Extended Absracts of Second International Conference of Mathematical and Numerical Aspects Wave Propagation*. SIAM, Delaware, U.S.A., 145-147.

55. Bulatov, V.V., Vladimirov, Y.V., Danilov,V.G. and Dobrokhotov, S.Y. 1994 On motion of the point algebraic singularity for 2D nonlinear hydrodynamical equations. *Math.Notices,* 3, 243-250.

56. Bulatov, V.V. and Vladimirov, Y.V. 1995 Calculation the internal gravity wave field associated with arbitrary unsteady motion of a source. *Izv.Akad.Nauk. Fluid Dynamics,* 3, 483-486.

57. Bulatov, V.V. and Vladimirov, Y.V. 1995 Internal gravity waves from an arbitrary moving sources. *Abstracts of Third International Conference on Mathematical and Numerical Aspects of Wave Propagation*. Mandelieu, France. SIAM-INRIA, Edited by G.Cohen, 783-784.

58. Bulatov,V.V., Vladimirov,Y.V., Danilov, V.G. and Dobrokhotov, S.Y. 1996 Calculations of hurricane trajectory on the basis of V.P.Maslov hypothesis. *Dokl.Akad.Nauk.*, 2, 6-11.

59. Bulatov, V.V. and Vladimirov, Y.V. 1998 Uniform far field asymptotic of internal gravity waves from a source moving in a stratified fluid layer with smoothly varying bottom. *Izv.Akad.Nauk. Fluid Dynamics,* 3, 388-395.

60. Bulatov, V.V. and Vladimirov, Y.V. 1998 Numerical simulations of internal gravity waves from an object moving in a stratified fluid on the basis of Green's function method. *Proceeding of the Fourth International Conference on Mathematical and Numerical Aspects of Wave Propagation*. SIAM, Golden, Colorado, U.S.A., 729-731.

61. Bulatov, V.V. and Vladimirov, Y.V. 1998 Propagation internal gravity waves in unsteady inhomogeneous stratified medium. *Proceeding of the Fourth International Conference on Mathematical and Numerical Aspects of Wave Propagation*. SIAM, Golden, Colorado, U.S.A., 784-786.

62. Bulatov, V.V. and Vladimirov, Y.V. 2000 Asymptotic of the critical regimes of internal wave generation. *Izv.Akad.Nauk. Fluid Dynamics,* 5, 734-737.

63. Bulatov, V.V., Vakorin, V.A. and Vladimirov, Y.V. 2004 Internal gravity waves in a stratified fluid with smoothly varying bottom. Cornell University, E-Print Archive, Paper ID: physics/0411016, http://arxiv.org/abs/physics/0411016.

64. Bulatov, V.V., Vakorin, V.A. and Vladimirov, Y.V. 2004 Weak singularity for two-dimensional nonlinear equations of hydrodynamics and propagation of shock waves. Cornell University, E-Print Archive, Paper ID: math-ph/0410058, http://arxiv.org/abs/math-ph/0410058.

65. Bulatov, V.V. and Vladimirov, Y.V. 2005 Internal gravity waves in inhomogeneous medium. Nauka, Moscow, pp.195 (in Russian).

66. Bulatov, V.V., Vakorin, V.A. and Vladimirov, Y.V. 2005 Critical regimes of internal gravity wave generation. Cornell University, E-Print Archive, Paper ID: math-ph/0511083, http://arxiv.org/abs/math-ph/0511083

67. Bulatov, V.V., Vakorin, V.A. and Vladimirov, Y.V. 2005 Internal gravity waves in stratified horizontally inhomogeneous media. Cornell University, E-Print Archive, Paper ID: math-ph/0511082, http://arxiv.org/abs/math-ph/0511082

68. Bulatov, V.V. and Vladimirov, Y.V. 2006 General problems of the internal gravity waves linear theory // Cornell University, 2006, E-Print Archive, Paper ID: physics/0609236, http://arxiv.org/abs/physics/0609236

69. Bulatov,V.V. and Vladimirov,Y.V. 2006 Dynamics of the internal gravity waves in the heterogeneous and nonstationary stratified mediums // Cornell University, 2006, E-Print Archive, Paper ID: physics/0611040, http://arxiv.org/abs/physics/0611040

70. Case, K. M. 1978 Some properties of internal waves. *Phys. Fluids*, 21(1), 18-29.

71. Castro, I. P. 1987 A note on lee wave structures in stratified flow over three-dimensional obstacles. *Tellus* A 39, 72-81.

72. Castro, I. P. and Snyder, W. H. 1988 Upstream motions in stratified flow. *J. Fluid Mech.* 187, 487-506.

73. Castro, I. P., Snyder, W. H. and Marsh, G. L. 1983 Stratified flow over three-dimensional ridges. *J. Fluid Mech*. 135, 261-282.

74. Cerasoli, C. P. 1978 Experiments on buoyant-parcel motion and the generation of internal gravity waves. *J. Fluid Mech*., 86(2), 247-271.

75. Chandrasekhar, S. 1961 Hydrodynamic and hydromagnetic stability. Oxford University Press, London, UK, pp.652

76. Chang, W. L. and Stevenson, T. N. 1975 Internal waves in a viscous atmosphere. *J. Fluid Mech*., 1975, 72(4), 773-786.

77. Chashechkin, Yu. D. 1989 Hydrodynamics of a sphere in a stratified fluid. *Izv.Akad.Nauk Fluid Dyn*. 24, 1-7.

78. Chashechkin, Yu. D. and Makarov, S. A. 1984 Time-varying internal waves. *Dokl. Earth Sci. Sect*. 276, 210-213.

79. Chaudhuri, K. S. 1969 Directional variation of shallow water waves due to arbitrary periodic surface pressure. *J. Phys. Soc. Japan*, 26(4), 1048-1055.

80. Chee-Seng, L. 1981 Water waves generated by an oscillatory surface pressure travelling at critical speed. *Wave motion*, 3(2), 159-179.

81. Chen, H.-H. 1977 On rectangular pressure distribution of oscillating strength moving over a free surface. *J. Ship. Res*., 21(1), 11-23.

82. Cheng, H. K., Hefazi, H. and Brown, S. N. 1984 Topographically generated cyclonic disturbance and lee waves in a stratified rotating fluid. *J. Fluid Mech*. 141, 431-453.

83. Chomaz, J.-M., Bonneton, P. and Hopfinger, E. J. 1993 The structure of the near wake of a sphere moving horizontally in a stratified fluid. *J. Fluid Mech*. (in press).

84. Chui, C.K. 1992 An introduction to wavelets. Academic Press, pp. 266.

85. Cimbala, J. M. and Park, W. J. 1990 An experimental investigation of the turbulent structure in a two-dimensional momentumless wake. *J. Fluid Mech*. 213, 479-509.

86. Clark, T. L. and Peltier, W. R. 1977 On the evolution and stability of finite-amplitude mountain waves. *J. Atmos. Sci*. 34, 1715-1730.

87. Cole, J. D. and Greifinger, C. 1969 Acoustic-gravity waves from an energy source at the ground in an isothermal atmosphere. *J. Geophys. Res*. 74, 3693-3703.

88. Gordeichik, B.N. and Ter-Krikorov, A.M. 1996 Uniform approximation of the fundamental solution of the equation of internal waves. *Appl.Mathem.Mech.*, 3, 439-447.

89. Crapper, G. D. 1959 A three-dimensional solution for waves in the lee of mountains. *J. Fluid Mech*. 6, 51-76.

90. Crapper, G. D. 1962 Waves in the lee of a mountain with elliptical contours. *Phil. Trans. R. Soc. Lond*. A 254, 601-623.

91. Crighton, D. G. and Oswell, J. E. 1991 Fluid loading with mean flow. I. Response of an elastic plate to localized excitation. *Phil. Trans. R. Soc. Lond*. A 335, 557-592.

92. Dagan, G. 1975 Waves and wave resistance of thin bodies moving at low speed: the free-surface nonlinear effect. *J. Fluid Mech*., 69(2), 405-416.

93. Dagan, G. and Miloh, T. 1982 Free-surface flow past oscillating singularities at resonant frequency. *J. Fluid Mech*., 120, 139-154.

94. Dahlqvist, H and Kallen, E. 1991 Field experiments to study internal wave generation in the Baltic. *FOA rapport C*20834-2.7. Mars 1991.

95. Debnath, L. 1968 On transient development of surface waves. *Z. angew. Math. und Phys*., 19(6), 948-961.

96. Debnath, L. 1969 On three dimensional transient wave motions on a running stream. *Meccanica*, 4(2), 122-128.

97. Debnath, L. 1969 On initial development of axisymmetric waves due to sources. *Indian J. Phys*., 43(11), 680-692.

98. Debnath, L. 1969 An asymptotic treatment of the transient development of .axisymmetric surface waves. *Appl. Sci. Res*., 21(1), 24-36.

99. Debnath, L. 1970 On transient development of ship waves on a running stratified ocean. *Meccanica*, 5(4), 277-284.

100. Debnath, L. 1983 Unsteady axisymmetric capillary-gravity waves in a viscous fluid. *Indian J. Pure and Appl. Math*., 14(4), 540-553.

101. Debnath, L., Bagehi, K. K. and Mukherjee, S. 1977 Capillary-gravity waves in a viscous fluid. *Acta mech*., 28(1-4), 313-319.

102. Debnath, L. and Rosenblat, S. 1969 The ultimate approach to the steady state in the generation of waves on a running stream. *Quart. J. Mech. and Appl. Math*., 22(2), 221-223.

103. Debnath, L. and Vann, J. M. 1974 Propagation of water waves with surface tension due to submerged sources. *Bull. Acad. pol. sci. Ser. sci. techn*., 22(11), 901-908.

104. Delisi, D. P. and Orlanski, I. 1975 On the role of density jumps in the reflexion and breaking of internal gravity waves. *J. Fluid Mech*., 69(3), 445-464.

105. Dickinson, R. E. 1969 Propagators of atmospheric motions. 1. Excitation by point impulses. *Rev. Geophys*. 7, 483-514.

106. Djordjevic, V. D. 1976 Einige Beispiele von hyperbolischen Schwingungen in geschichteten Flussigeiten. *Z. angew. Math. und Mech.*, 56(3), 174-176.

107. Doctors, L. J. 1978 Hydrodynamic power radiated by a heaving and pitching air-cushion vehicle. *J. Ship Res.*, 22(2), 67-79.

108. Dokuchaev, V. P. and Dolina, I. S. 1977 Radiation of internal waves by sources in an exponentially stratified fluid. *Izv. Atmos. Ocean. Phys.* 13, 444-449.

109. Dotsenko, S.F. 1991 Generation of long internal waves in the ocean by a moving pressure zone. *Phys.Oceanogr.,* 3, 163-170.

110. Drazin, P. G. 1961 On the steady flow of a fluid of variable density past an obstacle. *Tellus* 13, 239-251.

111. Droughton, J. V. and Chen, C. F. 1972 The channel flow of a density-stratified fluid about immersed bodies. *Trans. ASME: J. Basic Eng.*, 94(1), 122-130.

112. Dugan, J. P., Warn-Varnas, A. C. and Piacsek, S. A. 1976 Numerical results for laminar mixed region collapse in density stratified fluid. *Computers and Fluids*, 4(2), 109-121.

113. Eaiock, T. R. and Wu, G. X. 1986 Wave resistance and lift on cylinders by a coupled element technique. *Int. Shipbuild. Progr.*, 33(377), 2-9.

114. Eckart, C. 1961 Internal waves in the ocean. *Phys. Fluids*, 4(7), 791-799.

115. Ekman, V. W. 1904 On dead water. In: Norwegian North Polar Expedition, 1893-1896, Scientific Results (ed. by F. Nansen), Longmans (1906), 5(15).

116. Eliassen, A. and Palm, E. 1960 On the transfer of energy in stationary mountain waves. *Geophysica Norvegica*, 22, 1-23.

117. Engevik, L. 1975 On the indeterminancy of the problem of stratified fluid flow over a barrier and related problems. *Z. Angew. Math. und Phys.*, 26(6), 831-834.

118. Farell, C. 1973 On the wave resistance of a submerged spheroid. *J. Ship. Res.*, 17(1), 1-11.

119. Farell, C. and Guven O. 1973 On the experimental determination of the resistance components of a submerged spheroid. *J. Ship. Res.*, 17(2), 72-79.

120. Felsen, L. B. 1969 Transients in dispersive media, Part I: Theory. *IEEE Trans. Antennas Propag.* 17, 191-200.

121. Foldvik, A. and Wurtele, M. G. 1967 The computation of the transient gravity wave. *Geophys. J. R. Astron. Soc.* 13, 167-185.

122. Forbes, L. K. and Schwartz, L. W. 1982 Free-surface flow over a seircircular obstruction. *J. Fluid Mech*., 114, 299-314.

123. Fox, D. W. 1976 Transient solutions for stratified fluid flows. *J. Res. Nat. Bur. Stand*., B80(1), 79-88.

124. Freund, D. D. and Meyer, R. E. 1972 On the mechanism of blocking in a stratified fluid. *J. Fluid Mech*., 54(4), 719-744.

125. Garret, C.J.R. and Munk, W.H. 1979 Internal waves in the ocean. *Ann.Rev.Fluid Mech.*, 11, 339-369.

126. Gargett, A. 1989 Ocean turbulence. *Ann. Rev. Fluid Mech*., 21, 419-451.

127. Gärtner, U. 1983 A note on the visualization and measurement of the internal wave field behind a cylinder moving through a stratified fluid. Geophys. Astrophys. *Fluid Dyn*. 26, 139-145.

128. Gärtner, U. 1983 Visualization of particle displacement and flow in stratified salt water. *Exp. Fluids* 1, 55-56.

129. Gärtner, U., Wernekinck, U. and Merzkirch, W. 1986 Velocity measurements in the field of an internal gravity wave by means of speckle photography. *Exp. Fluids* 4, 283-287.

130. Gazdar, A. S. 1973 Generation of waves of small amplitude by on obstacle placed on the bottom of a running stream. *J. Phys. Soc. Japan*, 34(2), 530-538.

131. Gerkema, T. 2002 Application of an internal tide generation model to baroclinic spring-neap cycles. *J.Geophys.Res.,* 107, 1(7)-7(7).

132. Ghosh, B. and Chaudhuri, K. S. 1983 Waves due to a moving line impulse on the surface of running stream of finite depth. *Meccanica*, 18(1), 16-20.

133. Gill, A.E. 1982 Atmosphere-Ocean Dynamics. Academic

134. Gilreath, H. E. and Brandt, A. 1983 Experiments on the generation of internal waves in a stratified fluid. *AIAA Pap*., 1704, 12.

135. Gilreath, H. E. and Brandt, A. 1985 Experiments on the generation of internal waves in a stratified fluid. *AIAA J*. 23, 693-700.

136. Giovanangeli, J. P. and Memponteil, A. 1985 Resonant and non-resonant waves excited by periodic vortices in airflow over water. *J. Fluid Mech*., 159, 69-84.

137. Goodman, L. and Levine, E. R. 1977 Generation of oceanic internal waves by advecting atmospheric fields. *J. Geophys. Res*., 82(12), 1711-1717.

138. Gordon, D., Klement, U. R. and Stevenson, T. N. 1975 A viscous internal wave in a stratified fluid whose buoyancy frequency varies with altitude. *J. Fluid Mech.*, 69(3), 615-624.

139. Gordon, D. and Stevenson, T. N. 1972 Viscous effects in a vertically propagating internal wave. *J. Fluid Mech.*, 56, 629-639.

140. Gorgui, M. A. and Kassem, S. E. 1978 Basic singularities in the theory of internal waves. *Quart. J. Mech. and Appl. Math.*, 31(1), 31-48.

141. Gorodtsov, V. A. 1980 Radiation of internal waves during vertical motion of a body through a nonuniform liquid. *J. Engng Phys.* 39, 1062-1065.

142. Gorodtsov, V. A. 1981 Radiation of internal waves by rapidly moving sources in an exponentially stratified liquid. *Sov. Phys. Dokl.* 26, 229-230.

143. Gorodtsov, V. A. 1983 Evolution of axisymmetric vorticity distributions in an ideal incompressible stratified liquid. *Appl. Maths Mech.* 47, 479-484.

144. Gorodtsov, V. A. 1991 Collapse of asymmetric perturbations in a stratified fluid. *Izv.Akad.Nauk Fluid Dyn.* 26, 834-840.

145. Gorodtsov, V. A. 1991 High-speed asymptotic form of the wave resistance of bodies in a uniformly stratified liquid. *J. Appl. Mech. Tech. Phys.* 32, 331-337.

146. Gorodtsov, V. A. and Teodorovich, E. V. 1980 On the generation of internal waves in the presence of uniform straight-line motion of local and nonlocal sources. *Izv. Atmos. Ocean. Phys.* 16, 699-704.

147. Gorodtsov, V. A. and Teodorovich, E. V. 1981 Two-dimensional problem for internal waves generated by moving singular sources. *Izv.Akad.Nauk Fluid Dyn.* 16, 219-224.

148. Gorodtsov, V. A. and Teodorovich, E. V. 1982 Study of internal waves in the case of rapid horizontal motion of cylinders and spheres. *Izv.Akad.Nauk Fluid Dyn.* 17, 893-898.

149. Gorodtsov, V. A. and Teodorovich, E. V. 1983 Radiation of internal waves by periodically moving sources. *J. Appl. Mech. Tech. Phys.* 24, 521-526.

150. Gossard, E. E. and Hooke, W. H. 1978 Waves in the atmosphere. Amsterdam.

151. Gradstein, I.S. and Ryshik, I.M. 1980 Table of Integrals, Series and Products. Academic

152. Graevel, W. P. 1969 On the slow motion of bodies in stratified and rotating fluids. *Quart. J. .Mech. and Appl. Math.*, 22(1), 39-54.

153. Graham, E. W. 1973 Transient internal waves produced by a moving body in a tank of density-stratified fluid. *J. Fluid Mech.*, 61(3), 465-480.

154. Graham, E. W. and Graham, B. B. 1980 The tank wall effect on internal waves due to a transient vertical force moving at fixed depth in a density stratified fluid. *J. Fluid Mech.*, 97(1), 91-114.

155. Gray, E.P., Hart, R.W. and Farrel, R.A.. 1983 The structure of the internal wave Mach front generated by a point source moving in a stratified fluid. *Phys. Fluids*, 10, 2919-2931.

156. Grigor'ev, G. I. and Dokuchaev, V. P. 1970 On the theory of the radiation of acoustic-gravity waves by mass sources in a stratified isothermal atmosphere. *Izv. Atmos. Ocean. Phys.* 6, 398-402.

157. Grimshaw, R. H. J. 1969 Slow time-dependent motion of a hemisphere in a stratified fluid. *Mathematika* 16, 231-248.

158. Grue, J. and Palm, E. 1985 Wave radiation and wave diffraction from a submerged body in a uniform current. *J. Fluid Mech.*, 151, 257-278.

159. Han, T. Y., Meng, J. C. S. and Innis, G. E. 1983 An open boundary condition for incompressible stratified flows. *J. Comput. Phys.*, 49(2), 276-297.

160. Hanazaki, H. 1988 A numerical study of three-dimensional stratified flow past a sphere. J. Fluid Mech. 192, 393-419.

161. Hanazaki, H. 1989 Drag coefficient and upstream influence in three-dimensional stratified flow of finite depth. *Fluid Dyn. Res.* 4, 317-332.

162. Harband, J. 1976 Three-dimensional flow over a submerged object. *J. Eng. Math.*, 10(1), 1-21.

163. Hart, R. W. 1981 Generalized scalar potentials for linearized three-dimensional flows with vorticity. *Phys. Fluids* 24, 1418-1420.

164. Hartman, R. J. and Lewis, H. W. 1972 Wake collapse in a stratified fluid: linear treatment. *J. Fluid Mech.* 51, 613-618.

165. Haussling, H. J. 1977 Viscous flows of stably stratified fluids over barriers. *J. Atmos. Sci.* 34, 589-602.

166. Haussling, H. L and Van Eseltine, R. T. 1976 Unsteady air-cushion vehicle hydrodynamics using Fourier series. *J. Ship. Res.*, 20(2), 79-84.

167. Haussling, H. L and Van Eseltine, R. T. 1978 Waves and wave resistance for air-cushion vehicles with time-dependent cushion pressures. *J. Ship Res.*, 22(3), 170-177.

168. Hawthorne, W. R. and Martin, M. E. 1955 The effect of density gradient and shear on the flow over a hemisphere. *Proc. R. Soc. Lond.* A 232, 184-195.

169. Hearn, G. E. 1977 Alternative methods of evaluating Green's function in three-dimensional ship-wave problems. *J. Ship Res.*, 21(2), 89-93.

170. Hefazi, H. T. and Cheng, H. K. 1988 The evolution of cyclonic disturbances and lee waves over a topography in a rapidly rotating stratified flow. *J. Fluid Mech.* 195, 57-76.

171. Hendershott, M. C. 1969 Impulsively started oscillations in a rotating stratified fluid. *J. Fluid Mech.* 36, 513-527.

172. Higuchi, H. and T. Kubota, T. 1990 Axisymmetric wakes behind a slender body including zero-momentum configurations. *Phys. Fluids A* 2, 1615-1623.

173. Hinwood, J. B. 1972 The study of density-stratified flows up to 1945. Pt 2. Internal waves and interfacial effects. *Houille blanche*, 27(8), 709-722.

174. Hodges, B.R., Imberger, J., Saggio, A. and Winters, K. 2000 Modelling basin-scale internal waves in a stratified lake. *Limnol.Oceanogr.*, 45, 1603-1620.

175. Holschneider, M. 1995 Wavelets: an analysis tool. Clarendon Press, Oxford, pp. 426.

176. Hopfinger, E.J. 1987 Turbulence in stratified fluids: a review. *J.Geophys.Res.*, 92, 5287-5303.

177. Hopfinger, E. J., Flor, J.-B., Chomaz, J.-M. and Bonneton, P. 1991 Internal waves generated by a moving sphere and its wake in a stratified fluid. *Exp. Fluids* 11, 255-261.

178. Horn, D.A., Redekopp, L.G., Imberger, J. and Ivey, G.N. 2000 Internal wave evolution in a space-time varying field. *J.Fluid Mech.*, 424, 279-301.

*179. Hudimac, A. 1961 Ship waves in a stratified ocean.* J. Fluid Mech*., 11, 229-243.*

180. Hunt, J. C. R. and Richards, K. J. 1984 Stratified airflow over one or two hills. *Boundary-Layer Meteorol.*, 30(1-4), 223-259.

181. Hunt, J. C. R. and Snyder, W. H. 1980 Experiments on stably and neutrally stratified flow over a model three-dimensional hill. *J. Fluid Mech.* 96, 671-704.

182. Hurdis, D. A. and Pao, H.-P. 1976 Observations of wave motion and upstream influence in a stratified fluid. *Trans. ASME: J. Appl. Mech.*, 43(2), 22-226.

183. Hurley, D. G. 1969 The emission of internal waves by vibrating cylinders. *J. Fluid Mech.*, 36(4), 657-672.

184. Hurley, D. G. 1971 Aerofoil theory for a stratified fluid. *Quart. J. Mech. and Appl. Math.*, 24(1), 37-42.

185. Hurley, D. G. 1972 A general method for solving steady-state internal gravity wave problems. *J. Fluid Mech.* 56, 721-740.

186. Hyun, J. M. 1976 Internal wave dispersion in deep oceans calculated by means of two-variable expansion techniques. *J. Oceanogr. Soc. Jap.*, 32(1), 11-20.

187. Hyun, J. M. 1977 Internal wave dispersion in density-stratified deep oceans with a thermocline. *J. Oceanogr. Soc. Jap.*, 33(1), 16-22.

188. Jackson, J. D. 1975 Classical Electrodynamics (2nd edn). Wiley.

189. Janowitz, G. S. 1968 On wakes in stratified fluids. *J. Fluid Mech.*, 33(3), 417-432.

190. Janowitz, G. S. 1971 The slow transverse motion of bodies of a flat plate through non-diffusive stratified fluid. *J. Fluid Mech.*, 47(1), 171-181.

191. Janowitz, G. S. 1974 Line singularities in unbounded stratified flow. *J. Fluid Mech.*, 66(3), 455-464.

192. *Janowitz, G.S. 1981 Stratified flow over a bounded obstacle in a channel of finite height.* J. Fluid Mech.*, 110, 161-170.*

193. *Janowitz, G.S. 1984 Lee waves in a three-dimensional stratified flow.* J. Fluid Mech.*, 148, 97-108.*

194. *Johnston, T.M. and Merrifield, M.A. 2003 Internal tide scaterring at seamounts, ridges and islands.* J.Geophys.Res.*, 108, 1(11)-17(11).*

195. *Jones, W.L. 1969 Ray traicing for internal gravity waves.* J. Geoph. Res.*, 8, 2028-2033.*

196. *Jung-Tai Lin and Yih-Ho Pao 1979 Wakes in a stratified fluids.* Ann. Rev. Fluids Mech.*, 11, 317-338.*

197. *Kallen, E. 1987 Surface effects of vertically propagation waves in a stratified fluid.* J. Fluid Mech.*, 182, 111-125.*

198. Kamachi, M. and Honji, H. 1988 Interaction of interfacial and internal waves. *Fluid Dyn. Res.* 2, 229-241.

199. Kao, T. W. and Pao, H.-P. 1980 Wake collapse in the thermocline and internal solitary waves. *J. Fluid Mech.*, 97(1), 115-127.

200. Käse, R. H. 1971 Ober zweidimensionale luftdruckbedingte interne Wellen im exponentiell geschichteten Meer. *Dtsch. hydrogr. Z.*, 24(5), 193-209.

201. Kato, S. 1966 The response of an unbounded atmosphere to point disturbances. I. Time-harmonic disturbances. *Astrophys. J.* 143, 893-903.

202. Kato, S. 1966 The response of an unbounded atmosphere to point disturbances. II. Impulsive disturbances. *Astrophys. J.* 144, 326-336.

203. Kato, H. and Phillips, O.M. 1969 On the penetration of a turbulent layer into a stratified fluid. *J.Fluid Mech.*, 37, 643-655.

204. Keller, J.B. 1958 Surface waves on water of non-uniform depth. *J. Fluid Mech.*, 6, 607-614.

205. Keller, J.B., Levy, D.M. and Ahluwalia, D.S. 1981 Internal and surface wave production in a stratified fluid. *Wave Motion*, 3, 215-229.

206. Keller, J.B. and Mow, Van C. 1969 Internal wave propagation in a inhomogeneous fluid of non-uniform dept. *J. Fluid Mech.*, 2, 365-374.

207. Keller, J. B. and Munk, W. H. 1970 Internal wave wakes of a body moving in a stratified fluid. *Phys. Fluids*, 6, 1425-1431.

208. Klemp, J. B. and Lilly, D. K. 1978 Numerical simulation of hydrostatic mountain waves. *J. Atmos. Sci.* 35, 78-107.

209. Korving, C. 1981 A numerical method for the wave resistance of a moving pressure distribution on the free surface. *Lect. Notes Phys.*, 141, 254-259.

210. Kranzer, H. C. and Keller, J. B. 1959 Water waves produced by explosions. *J. Appl. Phys.*, 30(3), 398-407.

211. Krauss, W. 1966 Methoden und Ergebnisse der theoretischen Ozeanographie. Bd. 2. Interne Wellen. Berlin.

212. Krishna, D. V. and Sarma, L. V. K. V. 1969 Motion of an axisymmetric body in a rotating stratified fluid confined between two parallel planes. *J. Fluid Mech.* 38, 833-842.

213. Lai, R. Y. S. and Lee, C-M. 1981 Added mass of a spheroid oscillating: in a linearly stratified fluid. *Int. J. Eng. Sci.*, 19(11), 1411-1420.

214. Landau, L. and Lifchitz, E. 1970 Theorie des Champs. Mir.

215. Landau, L.D. and Lifshitz, E.M. 1987 Fluid Mechanics. Volume 6 in Course of Theoretical Physics, Pergamon press, Oxford, UK. Second edition, pp.539.

216. Larsen, L. H. 1969 Oscillations of a neutrally buoyant sphere in a stratified fluid. *Deep Sea Res*. 16, 587-603.

217. Laurent, L.T., Stringer, S., Garret, C. and Perraut-Joncas, D. 2003 The generation of internal tides at abrupt topography. *Deep-Sea Res.*, 50, 987-1003.

218. Laws, P. and Stevenson, T. N. 1972 Measurements of a laminar wake in a confined stratified fluid. *J. Fluid Mech.*, 54(4), 745-748.

219. Le Blond, P. H. and Mysak, L. A. 1978 Waves in the ocean. Amstepdam.

220.  Legg , S. 2004 Internal tides generated on a corrugated continental slope. Part 1: Cross-slope barotropic forcing. *J. Phys. Oceanography*, 34(1), 156-173.

221.  Liandrat, J. and Moret-Bailly, F. 1990 The wavelet transform: some applications to fluid dynamics and turbulence. *Europ. J. Mech., B/Fluids*, 1, 1-19.

222.  Lighthill, M. J. 1958 Introduction to Fourier Analysis and Generalised Functions. Cambridge University Press.

223.  Lighthill, M. J. 1960 Studies on magneto-hydrodynamic waves and other anisotropic wave motions. *Phil. Trans. R. Soc. Lond*. A 252, 397-430.

224.  Lighthill, M. J. 1965 Group velocity. *J. Inst. Maths Applies* 1, 1-28.

225.  Lighthill, M. J. 1967 On waves generated in dispersive systems by travelling forcing effects, with applications to the dynamics of rotating fluids. *J. Fluid Mech*. 27, 725-752.

226.  Lighthill, M. J. 1978 Waves in Fluids. Cambridge University Press.

227.  Lighthhill, M. J. 1986 An Informal Introduction to Theoretical Fluid Mechanics. Oxford University Press.

228.  Lin, J.-T. and Apelt, C. J. 1973 Stratified flow over a vertical barrier. *Lect. Notes Phys*., 19, 176-183.

229.  Lin, J.-T. and Pao, Y.-H. 1979 Wakes in stratified fluids. *Ann. Rev. Fluid Mech*. 11, 317-338.

230.  Lin, Q., Boyer, D. L. and Fernando, H. J. S. 1992 Turbulent wakes of linearly stratified flow past a sphere. *Phys. Fluids A* 4, 1687-1696.

231.  Lin, Q., Lindberg, W. R., Boyer, D. L. and Fernando, H. J. S. 1992 Stratified flow past a sphere. *J. Fluid Mech*. 240, 315-354.

232.  Liu, C. H. and Yeh, K. C. 1971 Excitation of acoustic-gravity waves in an isothermal atmosphere. *Tellus* 23, 150-163.

233.  Liu, J. T. C. 1979 Generation of interfacial gravity waves by submerged regions of fluctuating hydrodynamical motions and fluid inhomogeneities. *Phys. Fluids*, 22(5), 814-818.

234.  Lofquist, K. E. B. and Purtell, L. P. 1984 Drag on a sphere moving horizontally through a stratified liquid. *J. Fluid Mech*., 148, 271-284.

235.  MacKinnon, R. F., Mulley, R. and Warren F. W. G. 1969 Some calculations of. gravity wave resistance in an inviscid stratified fluid. *J. Fluid Mech*., 38(1), 61-73.

236.  Magnuson, A. H. 1977 The disturbance produced by an oscillatory pressure distribution in uniform translation on the surface of a liquid. *J. Eng. Math*., 11(2), 121-137.

237. Mahanti, N. C. 1976 Waves due to an oscillatory pressure on a stratified fluid: a uniformly valid asymptotic solution. *Appl. Sci. Res*. 32(2), 167-178.

238. Mahanti, N. C. 1977 Uniform asymptotic analysis of wave motion due to a periodic pressure on a stratified fluid of finite depth. *Quart. J. Mech. and Appl. Math.*, 30(4), 375-385.

239. Mahanti, N. C. 1977 Small amplitude surface waves due to an oscillatory pressure. *Isr. J. Technol.*, 15(6), 373-374.

240. Mahanti, N. C. 1978 Oscillation modes in superposed fluids. *Trans. ASME: J. Appl. Mech.*, 45(1), 204-205.

241. Mahanti, N. C. 1979 Small-amplitude internal waves due to an oscillatory pressure. *Quart. Appl. Math.*, 37(1), 92-94.

242. Mahanti, N. C. 1984 The Cauchy-Poisson problem for a nonhomogeneous. fluid of finite depth. *Proc. Indian Nat. Sci. Acad.*, 450(4), 327-334.

243. Makarov, S. A. and Chashechkin, Yu. D. 1981 Apparent internal waves in a fluid with exponential density distribution. *J. Appl. Mech. Tech. Phys*. 22, 772-779.

244. Makarov, S. A. and Chashechkin, Yu. D. 1982 Coupled internal waves in a viscous incompressible fluid. *Izv. Atmos. Ocean. Phys*. 18, 758-764.

245. Malvestuto, V. 1979 Internal wave motion in a periodic stratification. *Phys. Fluids*, 22(10), 1862-1867.

246. Manna, R. K. 1983 Forced oscillations in two-layer fluid of finite depth. *Trans. ASME: J. Appl. Mech.*, 50(3), 506-510.

247. Martin, S. and Long, R. R. 1968 The slow motion of a flat plate in a viscous stratified fluid. *J. Fluid Mech.*, 31(4), 669-688.

248. Maxworthy, T. 1980 On the formation of nonlinear internal waves from the gravitational collapse of mixed regions in two and three dimensions. *J. Fluid Mech.*, 96(1), 47-64.

249. McLaren, T. I., Pierce, A. D., Foul, T. and Murphy, B. L. 1973 An investigation of internal gravity waves generated by a buoyantly rising fluid in a stratified medium. *J. Fluid Mech*. 57, 229-240.

250. Mei, C. C. 1966 Surface wave pattern due to a submerged source travelling in a stratified ocean. *Rept. Hydrodynam. Lab. Mass. Inst. Technol.*, 92, 26.

251. Mei, C. C. 1969 Collapse of a homogeneous fluid in a stratified fluid. *Proc. 12th Int. Congr. Appl. Mech.*, Stanford Univ., 1968. Berlin, 521-330.

252. Mei, C. C. 1976 Flow around a thin body moving in shallow water. *J. Fluid Mech.*, 77(4), 737-751.

253. Meng, J.C.S. and Rottman, J.W. 1988 Linear internal waves generated by density and velocity perturbations in a linearly stratified fluid. *J. Fluid Mech.*, 186, 419-444

254. Miles, J. W. 1962 Transient gravity wave response to an oscillating pressure. *J. Fluid Mech.*, 13(1), 145-150.

255. Miles, J. M. 1968 Lee waves in stratified flow. Pt. 2. Semi-circular obstacle. *J. Fluid Mech.*, 33(4), 803-814.

256. Miles, J. W. 1969 The lee-wave regime for a slender body in a rotating flow. *J. Fluid Mech.* 36, 265-288.

257. Miles, J. W. 1969 Transient motion of a dipole in a rotating flow. *J. Fluid Mech.* 39, 433-442.

258. Miles, J. W. 1969 Waves and wave drag in stratified flows. *Proceedings of the XIIth International Congress of Applied Mechanics* (ed. by M. Hétényi & W. G. Vincenti), Springer, pp. 50-76.

259. Miles, J. W. 1971 Internal waves generated by a horizontally moving source. *Geophys. Fluid Dyn.* 2, 63-87.

260. Miles, J. W. 1972 Internal waves in a sheeted thermocline. *J. Fluid Mech.*, 53(3), 557-573.

261. Miles, J.W.and Chamberlain, P.G. 1998 Topographical scattering of gravity waves. *J. Fluid Mech.*, 361, 175-188

262. Miles, J. W. and Huppert, H. E. 1969 Lee waves in a stratified flow. Part 4. Perturbation approximations. *J. Fluid Mech.* 35, 497-525.

263. Miloh, T. and Dagan, G. 1985 A study of nonlinear wave resistance using integral equations in Fourier space. *J. Fluid Mech.*, 159, 433-458.

264. Miropol'skii, Yu. Z. 1978 Self-similar solutions of the Cauchy problem for internal waves in an unbounded fluid. *Izv. Atmos. Ocean. Phys.* 14, 673-679.

265. Miropol'skii, Yu. Z. and Shishkina, O.V. 2001 Dynamics of internal gravity waves in the ocean. Boston: Kluwer Academic Publishers.

266. Moore, D. W. and Spiegel, E. A. 1964 The generation and propagation of waves in a compressible atmosphere. *Astrophys. J.* 139, 48-71.

267. Morgan, G. W. 1953 Remarks on the problem of slow motions in a rotating fluid. *Proc. Camb. Phil. Soc.* 49, 362-364.

268.     Morozov, E.G., Vlasenko, V.I., Demidova, T.A. and Ledenev, V.V. 1999 Tidal internal wave propagation over large distance. *Oceanology*, 1,42-46.

269.     Morozov, E.G.2001 Several approaches to the investigation of tidal internal waves in the Northern part of the Pacific ocean. *Oceanology,* 2,171-175.

270.     Morozov, E.G., Trulsen, K., Velarde, M.G. and Vlasenko, V.I. 2002 Internal tides in the strait of Gibraltar. *J.Phys.Oceanogr.,* 32, 3193-3206.

271.     Morse, P. M. and Feshbach, H. 1953 Methods of Theoretical Physics. Part I. McGraw-Hill.

272.     Mowbray, D. E. and Rarity, B. S. H. 1967 A theoretical and experimental investigation of the phase configuration of internal waves of small amplitude in a density stratified liquid. *J. Fluid Mech*. 28, 1-16.

273.     Mowbray, D. E. and Rarity, B. S. H. 1967 The internal wave pattern produced by a sphere moving vertically in a density stratified liquid. *J. Fluid Mech*. 30, 489-495.

274.     Muller, P. and Liu, X. 2000 Scattering of internal waves at finite topography in two dimensions. Part I: theory and case studies.. J. Phys.Oceanogr., 30, 532-549.

275.     Muller, P. and Liu, X. 2000 Scattering of internal waves at finite topography in two dimensions. Part II: spectral calculations and boundary mixing. J. Phys.Oceanogr., 30, 550-563.

276.     Murdock, J. W. 1977 The near-field disturbance created by a body in a stratified medium with a free surface. *Trans. ASME J. Appl. Mech*. 44, 534-540.

277.     Nestegard, A. and Selavounos, P. D. 1984 A numerical solution of two-dimensional deep water wave-body problems. *J. Ship Res*., 28(1), 48-54.

278.     Newman, J. N. 1961 The damping of an oscillating ellipsoid near a free surface. *J. Ship Res*., 5(3), 44-58.

279.     Newman, J. N. 1977 Marine hydrodynamics. Cambridge: MIT Press.

280.     Newman, J. N. 1984 Double-precision evaluation of the oscillatory source potential. *J. Ship Res*., 28(3), 151-154.

281.     Newman, J. N. 1985 Algoritms for the free-surface Green function. *J. Eng. Math*., 19(1), 57-67.

282.     Newman, J. N., Sortland, B. and Vinje, T. 1984 Added mass and damping of rectangular bodies close to the free surface. *J. Ship Res*., 28(4), 216-225.

283.     Nguyen, T.-D., Fruman, D. and Luu, T. S. 1968 Sur une methode de calcul des ecoulements bidimensionnels avec surface libre autour des corps immerges. *S. r. Acad. sci*., 266(6), 382-385.

284. Noak, B.R. and Eckelmann, H. 1994 A low-dimensional Galerkin method for the three-dimensional flow around a circular cylinder. *Phys. Fluids*, 1, 124-142.

285. Noblesse, F. 1981 Alternative integral representations for the Green function of the theory of ship wave resistance. *J. Eng. Math.*, 15(4), 241-265.

286. Noblesse, F. 1982 The Green function in the theory of radiation and diffraction of regular water waves by a body. *J. Eng. Math.*, 16(2), 137-169.

287. Orlanski, I. and Ross, B. B. 1973 Numerical simulation of the generation and breaking of internal gravity waves. *J. Geophys. Res.*, 78(36), 8806-8826.

288. Peat, K. S. and Stevenson, T. N. 1975 Internal waves around a body moving in a compressible density-stratified fluid. *J. Fluid Mech.* 70, 673-688.

289. Peat, K. S. and Stevenson, T. N. 1976 The phase configuration of waves around a body moving in a rotating stratified fluid. *J. Fluid Mech.* 75, 647-656.

290. Pedlosky J. 2003 Waves in the ocean and atmosphere: introduction to wave dynamics. Academic.

291. Pellacani, C. 1976 Turning-level dynamics for infinitesimal disturbances in a stratified Boussinesq fluid. *Nuovo cim.*, B35(2), 333-341.

292. Pellacani, C. and Lupini, R. 1984 Bifurcations of the modal structure of internal waves in a fluid induced by a spatially periodical variation of the hydrostatic stability parameter. *Lett. nuovo cim.*, 40(4), 107-110.

293. Peltier, W. R. and Clark, T. L. 1979 The evolution and stability of finite-amplitude mountain waves. Part II: Surface wave drag and severe downslope windstorms. *J. Atmos. Sci.* 36, 1498-1529.

294. Peltier, W. R. and Clark, T. L. 1983 Nonlinear mountain waves in two and three spatial dimensions. *Q. J. R. Met. Soc.* 109, 527-548.

295. Peters, F. 1985 Schlieren interferometry applied to a gravity wave in a density-stratified liquid. *Exp. Fluids*, 3(5), 261-269.

296. Pezzoli, G. 1967 On the theory of emersion and impulse waves. A rigorous solution, with finite energy, of the Cauchy-Poisson problem. *Meccanica*, 2(1), 41-48.

297. Phillips, D. S. 1984 Analytical surface pressure and drag for linear hydrostatic flow over three-dimensional elliptical mountains. *J. Atmos. Sci.* 41, 1073-1084.

298. Phillips, O. M. 1977 The dynamics of the upper ocean. 2nd ed. Cambridge e. a.: Cambridge Univ. Press.

299. Pidcock, M. K. 1985 The calculation of Green's functions in three dimensional hydrodynamic gravity wave problems. *Int. J. Numer. Meth. Fluids*, 5(10), 891-909.

300. Pierce, A. D. 1963 Propagation of acoustic-gravity waves from a small source above the ground in an isothermal atmosphere. *J. Acoust. Soc. Am*. 35, 1798-1807.

*301. Pierce, A. D. 1981 Acoustics. An Introduction to its Physical Principles and Applications. McGraw-Hill.*

302. Pramanik, A. K. 1974 Waves due to a moving oscillatory surface pressure in a stratified fluid. *Trans. ASME: J. Appl. Mech*., 41(3), 571-574.

303. Pramanik, A. K. 1979 The effect of initial acceleration on the waves produced by a moving oscillatory surface pressure. *Meccanica*, 14(3), 145-150.

304. Pramanik, A. K. and Majumdar, S. R. 1985 Capillary-gravity waves generated in a viscous fluid. *Phys. Fluids*, 28(1), 46-51.

305. Pramanik, A. K. and Majumdar, S. R. 1985 Small-amplitude free-surface waves generated by moving oscillatory disturbances. *J. Fluid Mech*., 145, 405-415.

306. Queney, P. 1977 Synthese des travauix theoriques sur les perturbations de relief. *Meteorologie*, 8, 113-143; Pt. 2. *Meteorologie*, 9, 111-163.

307. Ramachandra Rao, A. 1973 A note on the application of a radiation condition for a source in a rotating stratified fluid. *J. Fluid Mech*. 58, 161-164.

308. Ramachandra Rao, A. 1975 On an oscillatory point force in a rotating stratified fluid. *J. Fluid Mech*. 72, 353-362.

309. Ramirez, C. and Renouard, D. 1998 Generation of internal waves over shelf. *Dyn. Atmosph. Oceans*, 2, 107-125.

310. Rao, A. R. 1979 Green's function solution of a water wave problem in a stratified ocean of finite depth. *Int. J. Eng. Sci*., 17(5), 527-532.

311. Rao, A. R. and Balan, K. C. 1977 Effect of viscosity on internal waves from a source in a wall. *Proc. Indian Acad. Sci*., A85(5), 351-366.

312. Rarity, B. S. H. 1967 The two-dimensional wave pattern produced by a disturbance moving in an arbitrary direction in a density stratified liquid. *J. Fluid. Mech*., 30(2), 329-335.

313. Redekopp, L. G. 1975 Wave patterns generated by disturbances travelling horizontally in rotating stratified fluids. *Geophys. Fluid Dyn*. 6, 289-313.

314. Rehm, R. G. 1976 A survey of selected aspects of stratified and rotating: fluids. *J. Res. Nat. Bur. Stand*., B80(3), 353-402.

315. Rehm, R. G. and Radt, H. S. 1975 Internal waves generated by a translating oscillating body. *J. Fluid Mech*. 68, 235-258.

316. Rhodes-Robinson, P. F. 1970 Fundamental singularities in a theory of water waves with surface tension. *Bull. Austral. Math. Soc*., 11(3), 317-333.

317. Rhodes-Robinson, P. F. 1980 On waves at an interface between two liquids. *Math. Proc. Cambridge Phil. Soc*., 88(1), 183-193.

318. Robertson, G. E., Seinfeld, J. H. and Leal, L. G. 1976 Wakes in stratified flow past a hot or cold two-dimensional body. *J. Fluid Mech*., 75(2), 233-256.

319. Rotunno, R. and Smolarkiewicz, P. K. 1991 Further results on lee vortices in low-Froude-number flow. *J. Atmos. Sci*. 48, 2204-2211.

320. Row, R. V. 1967 Acoustic-gravity waves in the upper atmosphere due to a nuclear detonation and an earthquake. *J. Geophys. Res*. 72, 1599-1610.

321. Sahin, I. and Magnuson, A. H. 1984 A numerical method for the solution of a line source under a free surface. *Ocean Eng*., 11(5), 451-461.

322. Salvesen, N. and Kerczek, C. 1976 Comparison of numerical and perturbation solutions of two-dimensional non-linear water-wave problems. *J. Ship Res*., 20(3), 160-170.

323. Sarma, L. V. K. V. and Krishna, D. V. 1972 Motion of a sphere in a stratified fluid. *Zastosow. Matem*. 13, 123-130.

324. Sarma, L. V. K. V. and Krishna, D. V. 1972 Oscillation of axisymmetric bodies in a stratified fluid. *Zastosow. Matem*. 13, 109-121.

325. Sarma, L. V. K. V. and Naidu, K. B. 1972a Source in a rotating stratified fluid. *Acta Mech*. 13, 21-29.

326. Sarma, L. V. K. V. and Naidu, K. B. 1972b Closed form solution for a point force in a rotating stratified fluid. *Pure Appl. Geophys*. 99, 156-168.

327. Schooley, A. A. and Hughes, B. A. 1972 An experimental and theoretical study of internal waves generated by the collapse of a two-dimensional mixed region in a density gradient. *J. Fluid Mech*., 51(1), 159-175.

328. Scorer, R. S. 1956 Airflow over an isolated hill. *Q. J. R. Met. Soc*. 82, 75-81.

329. Scorer, R. S. 1978 Environmental aerodynamics. New York e. a.

330. Sekerzh-Zen'kovich, S. Ya. 1979 A fundamental solution of the internal-wave operator. *Sov. Phys. Dokl*. 24, 347-349.

331. Sekerzh-Zen'kovich, S. Ya. 1981 Construction of the fundamental solution for the operator of internal waves. *Appl. Maths Mech*. 45, 192-198.

332. Sekerzh-Zen'kovich, S. Ya. 1982 Cauchy problem for equations of internal waves. *Appl. Maths Mech*. 46, 758-764.

333. Sen, A. R. 1962 Deep-water surface waves due to arbitrary periodic pressures. *Proc. Nat. Inst. Sci. India*, A28(4), 612-631.

334. Sharman, R. D. and Wurtele, M. G. 1983 Ship waves and lee waves. *J. Atmos. Sci*. 40, 396-427.

*335.    Shen, H.-T. and Farell, C. 1977 Numerical calculation of the wave integrals in the linearized theory of water waves.* J. Ship Res*., 21(1), 1-10.*

*336.    Shivamoggi, B. K. 1982 Asymptotic behaviour of water waves generated by an initial displacement over a localised region and the final period of diffusion of a localised vorticity distribution.* Rev. roum. sci. techn. Ser. mec. appl*., 27(1), 101-103.*

337. Simon, M, J. and Ursell, F. 1984 Uniqueness in linearized two-dimensional water-wave problems. *J. Fluid Mech*., 148, 137-154.

338. Simpson, J. E. 1982 Gravity currents in the laboratory, atmosphere, and ocean. *Annu. Rev. Fluid Mech*. Vol. 14. Palo Alto, Calif., 213-234.

339. Small, J. and Martin, J. 2002 The generation of non-linear internal waves in the Gulf of Oman. *Continental Shelf Res.*, 22, 1153-1182.

340. Smith, R. B. 1979 The influence of mountains on the atmosphere. In: *Adv. Geophys*. Vol. 21. New York e. a., 87-230.

341. Smith, R. B. 1980 Linear theory of stratified hydrostatic flow past an isolated mountain. *Tellus* 32, 348-364.

342. Smith, R. B. 1988 Linear theory of stratified flow past an isolated mountain in isosteric coordinates. *J. Atmos. Sci*. 45, 3889-3896.

343. Smith, R. B. 1989 Mountain-induced stagnation points in hydrostatic flow. *Tellus* A 41, 270-274.

344. Smith, S. H. 1965 Surface waves over a submerged circular cylinder. *Mathematika*, 12(2), 235-245.

345. Smolarkiewicz, P. K. and Rotunno, R. 1989 Low Froude number flow past three-dimensional obstacles. Part I: Baroclinically generated lee vortices. *J. Atmos. Sci*. 46, 1154-1164.

346. Snyder, W. H., Thompson, R. S., Eskridge, R. E., Lawson, R. E., Castro, I. P., Lee, J. T., Hunt, J. C. R. and Ogawa, Y. 1985 The structure of strongly stratified flow over hills: dividing-streamline concept. *J. Fluid Mech*., 152, 249-288.

347.	Sobolev, S. L. 1965 Sur une classe des problemes de physique mathematique. *Simposio Internazionale Sulle Applicazioni Dell'Analisi Alia Fisica Matematica*, Cagliari-Sassari, September 28-October 4 1964, Edizioni Cremonese (Roma), pp. 192-208.

348.	Soh, W. K. 1984 A numerical method for non-linear water waves. *Comput. and Fluids*, 12(2), 133-143.

349.	Srokosz, M. A. 1979 The submerged sphere as an absorber of wave power. *J. Fluid Mech.*, 95(4), 717-741.

350.	Staquet, C. and Sommeria, J. 2002 Internal gravity wave: from instabilities to turbulence. *Ann. Rev Fluid Mech.*, 34, 559-593.

351.	Stevenson, T. N. 1968 Some two-dimensional internal waves in a stratified fluid. *J. Fluid Mech*. 33, 715-720.

352.	Stevenson, T. N. 1969 Axisymmetric internal waves generated by a travelling oscillating body. *J. Fluid Mech*. 35, 219-224.

353.	Stevenson, T. N. 1973 The phase configuration of internal waves around a body moving in a density stratified fluid. *J. Fluid Mech*. 60, 759-767.

354.	Stevenson, T. N., Bearon, J. N. and Thomas, N. H. 1974 An internal wave in a viscous heat-conducting isothermal atmosphere. *J. Fluid Mech.*, 65(2), 315-323.

355.	Stevenson, T. N., Chang, W. L. and Laws, P. 1979 Viscous effects in lee waves. *Geophys. Astrophys. Fluid Dyn*. 13, 141-151.

356.	Stevenson, T. N. and Thomas, N. H. 1969 Two-dimensional internal waves generated by a travelling oscillating cylinder. *J. Fluid Mech*. 36, 505-511.

357.	Stevenson, T. N., Woodhead, T. J. and Kanellopulos, D. 1983 Viscous effects in some internal waves. *Appl. Sci. Res*, 40, 185-197.

358.	Stewartson, K. 1952 On the slow motion of a sphere along the axis of a rotating fluid. *Proc. Camb. Phil. Soc*. 48, 168-177.

359.	Sturova, I. V. 1974 Wave motions produced in a stratified liquid from flow past a submerged body. *J. Appl. Mech. Tech. Phys*. 15, 796-805.

360.	Sturova, I. V. 1978 Internal waves generated by local disturbances in a linearly stratified liquid of finite depth. *J. Appl. Mech. Tech. Phys*. 19, 330-336.

361.	Sturova, I. V. 1980 Internal waves generated in an exponentially stratified fluid by an arbitrarily moving source. *Izv.Akad.Nauk Fluid Dyn*. 15, 378-383.

362.	Subba Rao, V. and Prabhakara Rao, G. V. 1971 On waves generated in rotating stratified liquids by travelling forcing effects. *J. Fluid Mech*. 46, 447-464.

363. Suzuki, M. and Kuwahara, K. 1992 Stratified flow past a bell-shaped hill. *Fluid Dyn. Res*. 9, 1-18.

364. Sykes, R. I. 1978 Stratification effects in boundary layer flow over hills. *Proc. R. Soc. Lond*. A 361, 225-243.

365. Sysoeva, E. Ya. and Chashechkin, Yu. D. 1986 Vortex structure of a wake behind a sphere in a stratified fluid. *J. Appl. Mech. Tech. Phys*. 27, 190-196.

366. Sysoeva, E. Ya. and Chashechkin, Yu. D. 1991 Vortex systems in the stratified wake of a sphere. *Izv.Akad.Nauk Fluid Dyn*. 26, 544-551.

367. Tayler, A. B. and Driessche, P. Van Der 1974 Small amplitude surface waves due to a moving source. *Quart. J. Mech. and Appl. Math*., 27(3), 317-345.

368. Teodorovich, E. V. and Gorodtsov, V. A. 1980 On some singular solutions of internal wave equations. *Izv. Atmos. Ocean. Phys*. 16, 551-553.

369. Thomas, N. H. and Stevenson, T. N. 1972 A similarity solution for viscous internal waves. *J. Fluid Mech*., 54(3), 495-506.

370. Thomas, N. H. and Stevenson, T. N. 1973 An internal wave in a viscous ocean stratified by both salt and heat. *J. Fluid Mech*., 61(2), 301-304.

371. Thorpe, S.A. 1975 The excitation, dissipation and interaction of internal waves in deep ocean. *J. Geoph. Res*., 3, 329-338.

372. Thorpe, S.A.. 1999 On the breaking of internal waves in the ocean. *J. Phys. Oceanogr*., 29, 2433-2441.

373. Tolstoy, I. 1973 Wave Propagation. McGraw-Hill.

374. Trubnikov, B. N. 1959 The three-dimensional problem of the flow over a barrier of an air current unbounded at the top. *Dokl. Earth Sci. Sect*. 129, 1136-1138.

375. Trustrum, K. 1971 An Oseen model of the two-dimensional flow of a stratified fluid over an obstacle. *J. Fluid Mech*., 50(1), 177-188.

376. Turner, J.S. 1973 Buoyancy effects in fluids. Cambridge University Press, London, UK, pp.367

377. Umeki, M. and Kambe, T. 1989 Stream patterns of an isothermal atmosphere over an isolated mountain. *Fluid Dyn. Res*. 5, 91-109.

378. Van Dyke, P. 1973 Bottom depth effects on regular surface waves due to a submerged Rankine body with attached vertical column. *J. Hydronautics*, 7(2), 78-84.

379. Vladimirov, V. A. and Il'in, K. I. 1991 Slow motions of a solid in a continuously stratified fluid. *J. Appl. Mech. Tech. Phys*. 32, 194-200.

380. Vladimirov, Y.V. 1989 Internal wave field in the neighborhood of a front excited by a source moving over smoothly varying bottom. *Appl.Mech.Techn.Phys.*, 4, 592-597.

381. Voisin, B. 1991 Internal wave generation in uniformly stratified fluid. Part 1. Green's function and point sources. *J. Fluid Mech.*, 231, 439-480.

382. Voisin, B. 1991 Rayonnement des ondes internes de gravite. Application aux corps en mouvement. Ph.D. thesis, Universite Pierre et Marie Curie.

383. Voisin, B. 1994 Internal wave generation in uniformly stratified fluids. Part II. Moving point sources. *J. Fluid Mech.*, 261, 333-374.

384. Warren, F. W. G. 1960 Wave resistance to vertical motion in a stratified fluid. *J. Fluid Mech*. 7, 209-229.

385. Warren, F. W. G. 1961 The generation of wave energy at a fluid interface by the passage of a vertically moving slender body. *Quart. J. Roy. Meteorol. Soc.*, 87(371), 43-54.

386. Watson, G. N. 1966 A Treatise on the Theory of Bessel Functions (2nd edn). Cambridge University Press.

387. Wehausen, J. V. 1973 The wave resistance of ships. In: *Adv. Appl. Mech.*, 13, 93-245.

388. Wehausen, J. V. and Laitone, E. V. 1960 Surface waves. In: *Handbuch der Physik*. Vol. 9. Berlin: Springer Verlag, 446-778.

389. Wei, S. N., Kao, T. W. and Pao, H. P. 1975 Experimental study of upstream influence in the two-dimensional flow a stratified fluid over an obstacle. *Geophys. Fluid Dyn.*, 6, 315-336.

390. Whitham, G. B. 1974 Linear and nonlinear waves. New York.

391. Wickerts, S. and Kallen, E. 1990 Observations and modeling on internal waves in a strongly stratified ocean. *FOA rapport C*20784-2.7, Februari 1990.

392. Winters, K.B., Lombard, P.N., Riley, J.J. and D'Asaro, E.A. 1995 Available potential energy and mixing in density-stratified fluids. *J.Fluid Mech.*, 289, 115-124.

393. Wolfe, P. and Lewis, R.M. 1966 Progressing waves radiated from a moving point source in an inhomogeneous media. *J. Diff. Equat.*, 3, 328-350.

394. Wong, K. K. and Kao, T. W. 1970 Stratified flow over extended obstacles and its application to topographical effect on vertical wind shear. *J. Atmos. Sci.*, 27(6), 884-889.

395. Woodhead, T. J. 1983 The phase configuration of the waves around an accelerating disturbance in a rotating stratified fluid. *Wave Motion* 5, 157-165.

396. Wu, J. 1969 Mixed region collapse with internal wave generation in a density-stratified medium. *J. Fluid Mech*. 35, 531-544.

397. Wu, T. Y.-T. 1965 Three-dimensional internal gravity waves in a stratified free-surface flow. *Z. Angew. Math. Mech. Send*. 45, T194-T195.

398. Wu, T. Y.-T. and Mei, C. C. 1967 Two-dimensional gravity waves in a stratified ocean. *Phys. Fluids*, 10(3), 482-486.

399. Wunsch, C. 1968 On the propagation of internal waves up a slope. *Deep-Sea Res*., 15, 251-258.

400. Wurtele, M. G. 1957 The three-dimensional lee wave. *Beitr. Phys. Atmos*. 29, 242-252.

401. Yanovitch, M. 1962 Gravity waves in a heterogeneous incompressible fluid. *Comm. Pure Appl. Math*., 1, 45-61.

402. Yuen, H.C. and Lake, B.M. 1982 Nonlinear dynamics of deep-water gravity waves. *Advanced in Applied Mechanics,* 22, 67-229.

403. Yeung, R. W. 1975 Surface waves due to a maneuvering air-cushion vehicle. *J. Ship Res*., 19(4), 224-242.

404. Yeung, R. W. 1982 Numerical methods in free-surface flows. In: *Annu. Rev. Fluid Mech*., Vol. 14. Palo Alto, Calif., 395-442.

405. Yeung, R.W., Rhines, P.B. and Garrett, C.J.R. 1982 Shear-Flow dispersion, internal waves and horizontal mixing in the ocean. *J. Phys. Oceanogr*., 12, 515-527.

406. Yih, C.-S. 1980 Stratified flows. New York: Acad. Press.

407. Zavol'skii, N. A. and Zaitsev, A. A. 1984 Development of internal waves generated by a concentrated pulse source in an infinite uniformly stratified fluid. *J. Appl. Mech. Tech. Phys*. 25, 862-867.

408. Zwick, W. 1964 Nichtlineare Methode zur Berechnung des Einflusses der freien Oberflache auf die Umstromung eines elliptischen Zylinders. *Monatsber. Dtsch. Akad. Wiss*. Berlin, 6(12), 894-901.

409. Zwick, W. and Hebermehl, G. 1967 Numerische Ergebnisse bei der Berechnung des Einflusses einer freien Oberflache auf die Umstromung eines Zylinder. *Monatsber. Dtsch. Akad. Wiss*. Berlin, 9(6-7), 396-404.

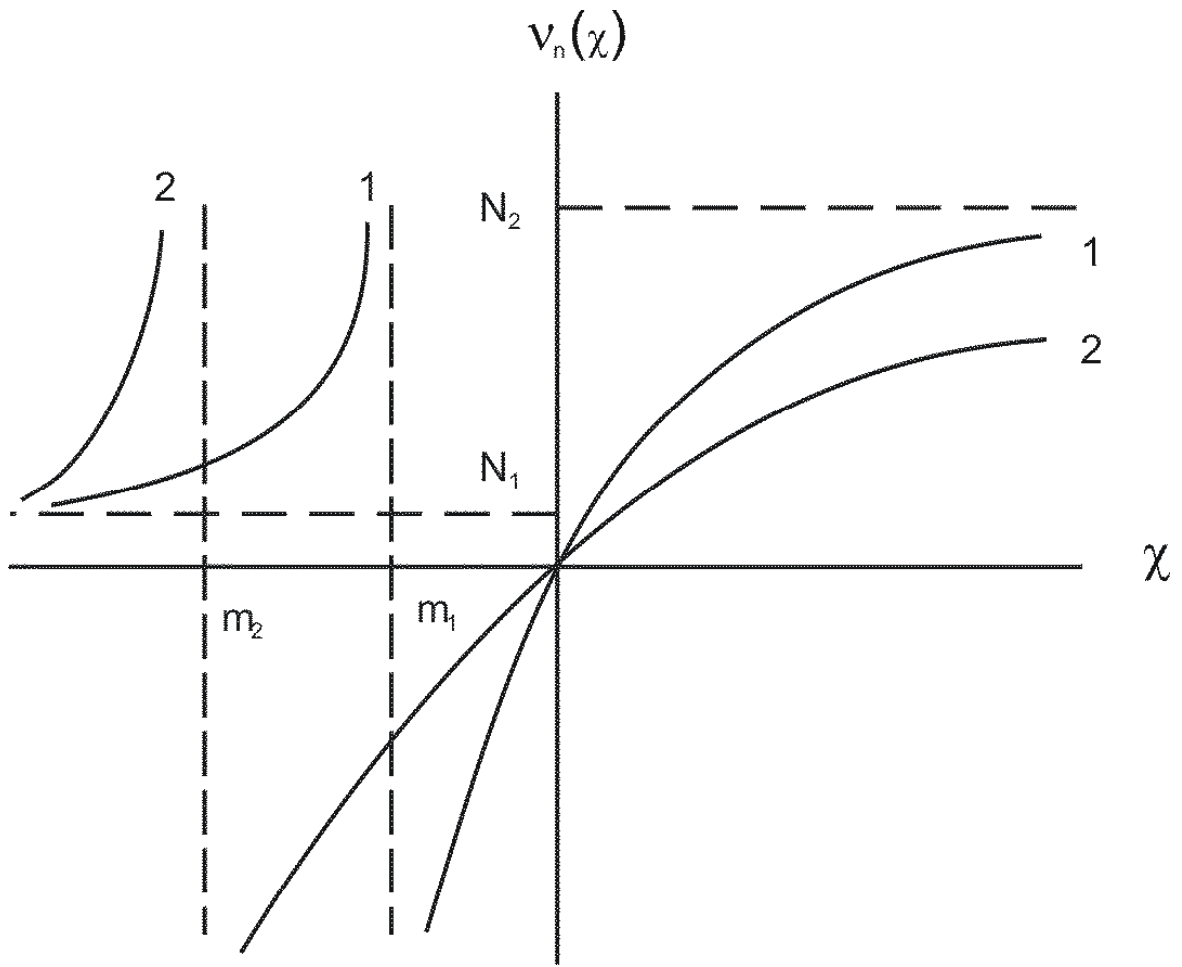

Fig. B1  Qualitative behavior of the dispersive curves $v_n(\chi)$ for the first two modes (the curve 1 - the first mode, 2 - the second mode,), $N_1$ - the minimal value of Brunt-Väisälä frequency $N(z)$, $N_2$ - the maximum value of $N(z)$, $m_n = -\dfrac{\pi^2 n^2}{H^2}$ ($n=1,2$).

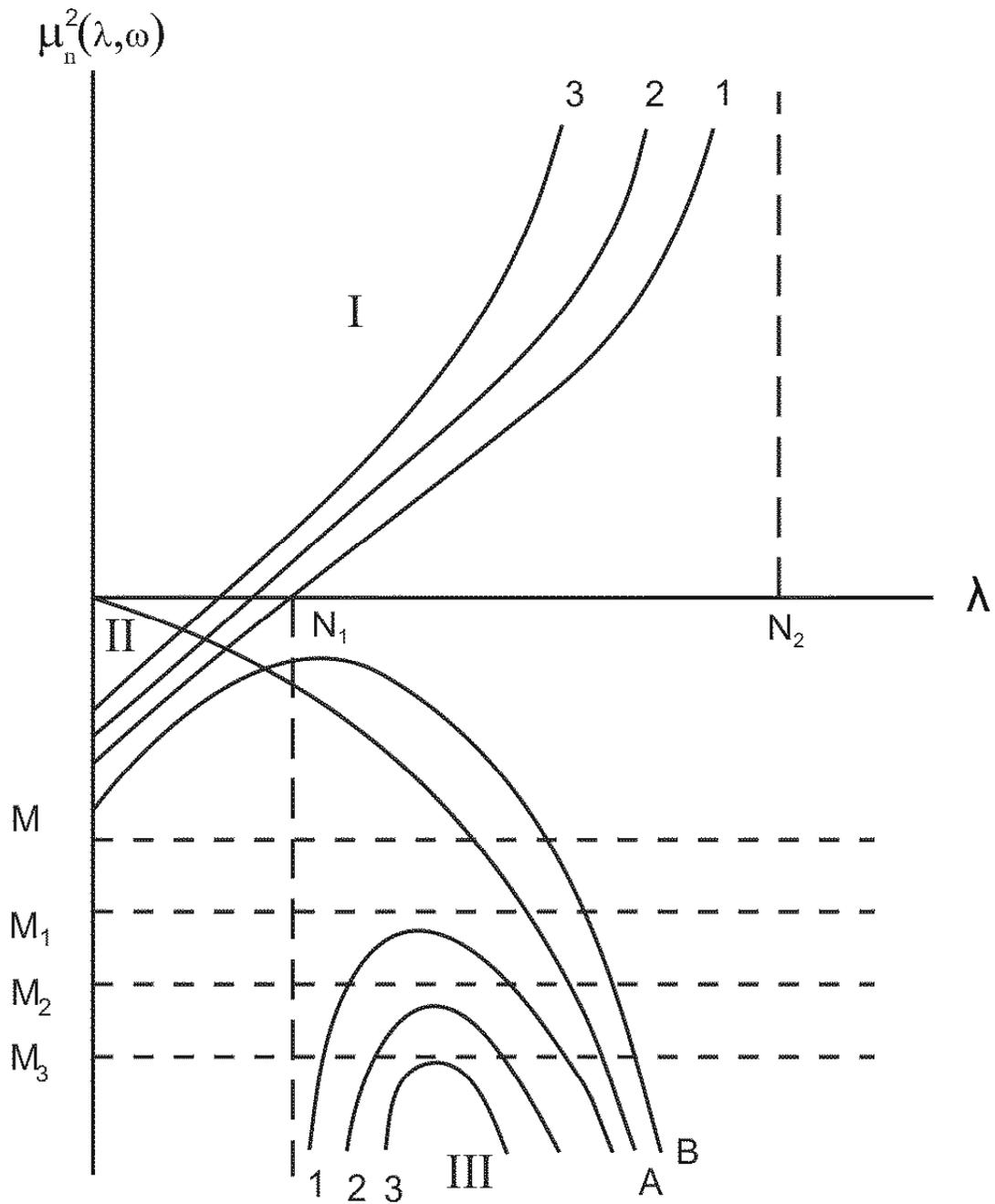

Fig. B2  Qualitative behavior of the dispersive curves $\mu_n^2(\lambda,\omega)$ for the first three modes at $\omega > 0$ (the curve 1 - the first mode, 2 - the second mode, 3 - the third mode), the curve A – function $(-\lambda^2)$, the curve B – function $-(\lambda-\omega/V)^2$, $N_1$ - the minimal value of Brunt-Väisälä frequency $N(z)$, $N_2$ - the maximum value of $N(z)$, $M = \omega^2 V^{-2}$, $M_n = -\left(\dfrac{\pi n}{\widetilde{H}(\infty)}\right)^2$.

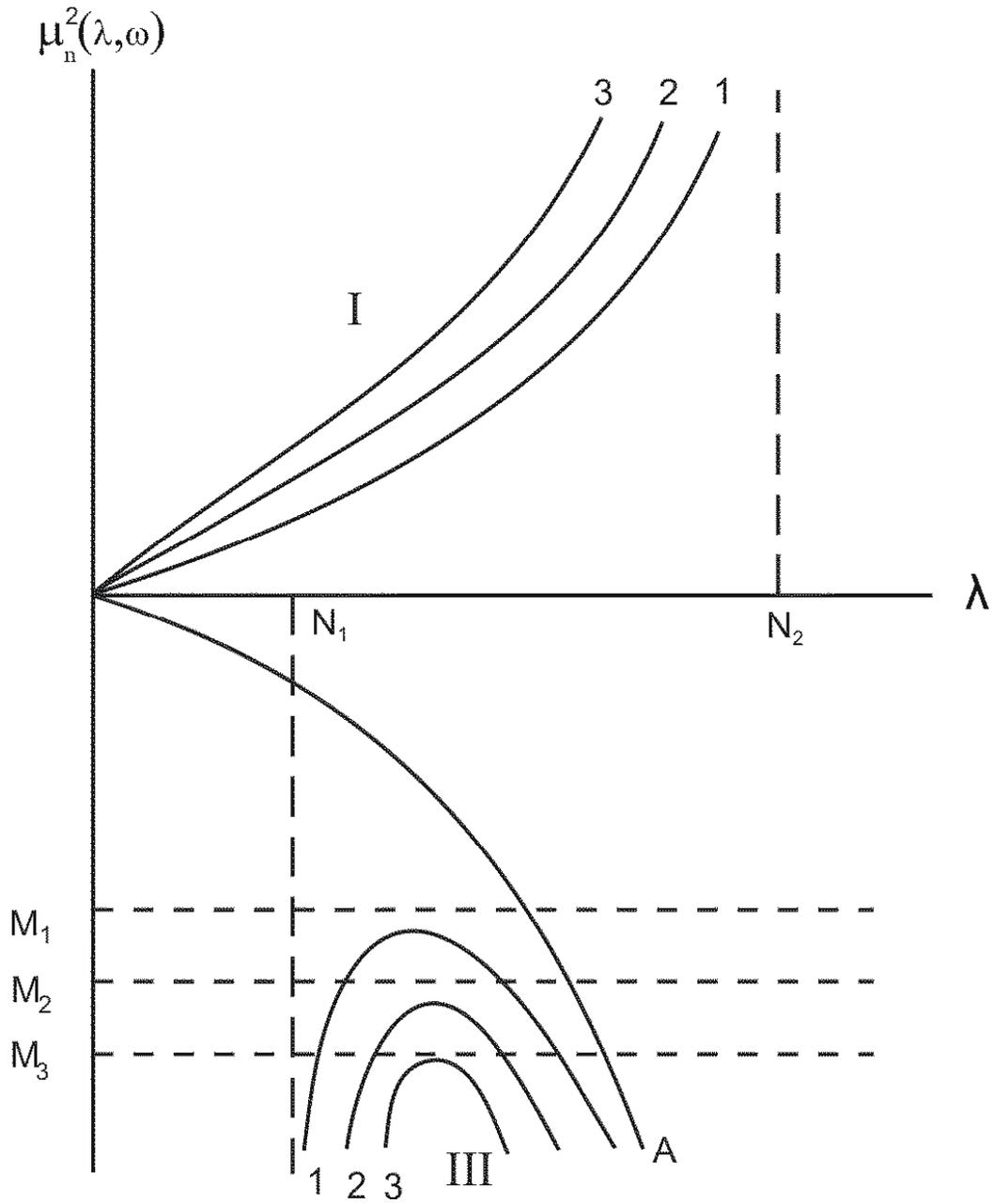

Fig. B3   Qualitative behavior of the dispersive curves $\mu_n^2(\lambda,\omega)$ for the first three modes at $\omega = 0$ (the curve 1 - the first mode, 2 - the second mode, 3 - the third mode), the curve A – function $(-\lambda^2)$, the curve B – function $-(\lambda-\omega/V)^2$, $N_1$ - the minimal value of Brunt-Väisälä frequency $N(z)$, $N_2$ - the maximum value of $N(z)$, $M = \omega^2 V^{-2}$, $M_n = -\left(\dfrac{\pi n}{\widetilde{H}(\infty)}\right)^2$.

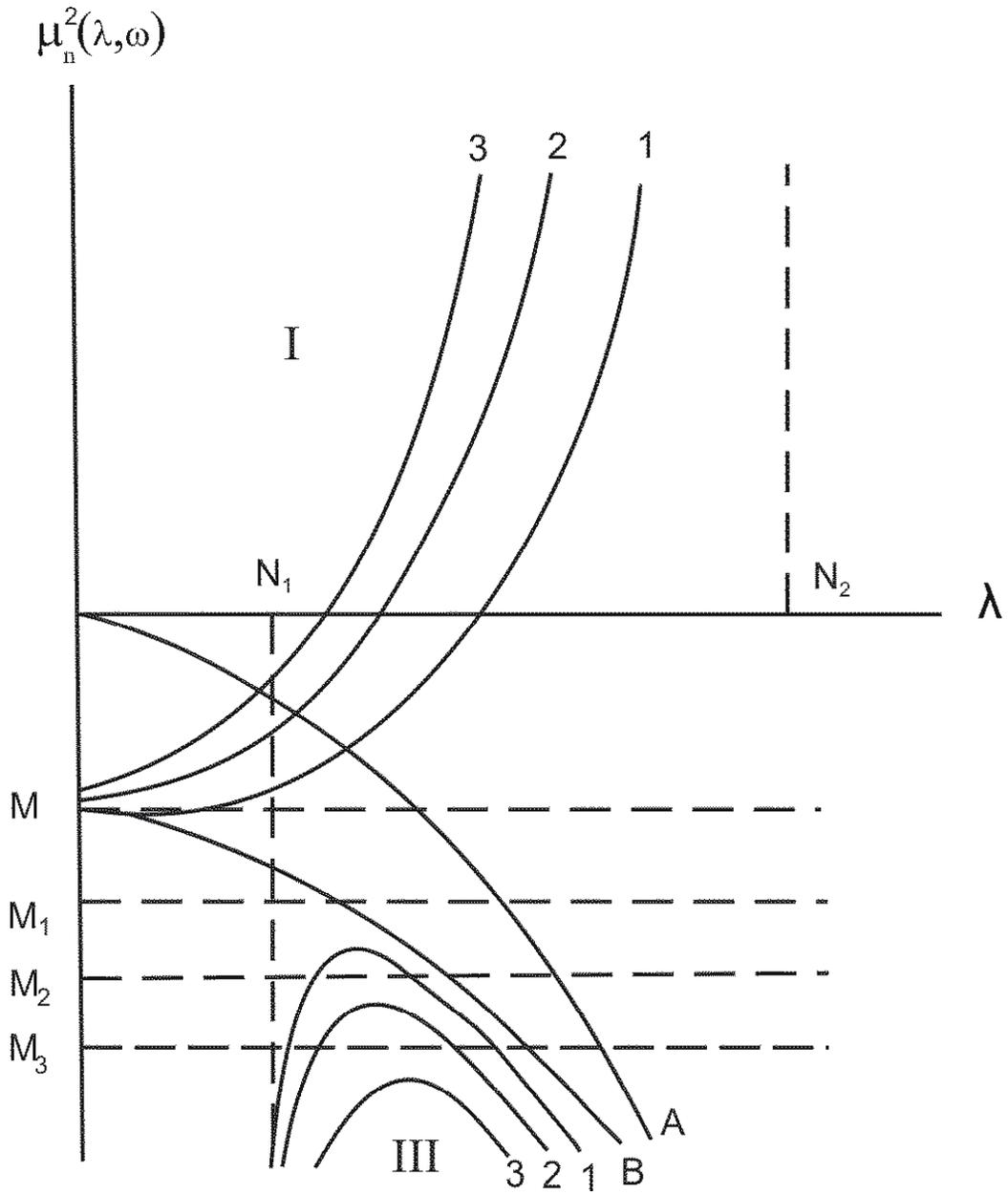

Fig. B4  Qualitative behavior of the dispersive curves $\mu_n^2(\lambda,\omega)$ for the first three modes at $\omega < 0$ (the curve 1 - the first mode, 2 - the second mode, 3 - the third mode), the curve A – function $(-\lambda^2)$, the curve B – function $-(\lambda-\omega/V)^2$, $N_1$ - the minimal value of Brunt-Väisälä frequency $N(z)$, $N_2$ - the maximum value of $N(z)$, $M = \omega^2 V^{-2}$, $M_n = -\left(\dfrac{\pi n}{\widetilde{H}(\infty)}\right)^2$.